\documentclass[%
aps,
 pra,%
 amsmath,amssymb,
 reprint,%
 superscriptaddress,
]{revtex4-1}

\usepackage{dcolumn}
\usepackage{graphicx}
\graphicspath{{figure/}}
\usepackage{lineno}
\usepackage{bm} %
\usepackage[pdftex,colorlinks=true,bookmarks=false,citecolor=blue,urlcolor=blue]{hyperref}
\usepackage{upgreek}
\usepackage{xcolor}

\usepackage{braket}
\newcommand{\scE}{\mathcal{E}}
\newcommand*\diff{\mathop{}\!\mathrm{d}}
\newcommand*\Diff[1]{\mathop{}\!\mathrm{d^#1}}
\newcommand{\olr}[1]{\overset{\text{\tiny$\leftrightarrow$}}{#1}}

\usepackage{titlesec} 

\begin{document}

\title{\vspace{-15mm}\fontsize{19pt}{10pt}\selectfont\textbf{High-speed tunable microwave-rate soliton microcomb}} 

\author{Yang He}
\thanks{These two authors contributed equally.}
\affiliation{Department of Electrical and Computer Engineering, University of Rochester, Rochester, NY 14627}

\author{Raymond Lopez-Rios}
\thanks{These two authors contributed equally.}
\affiliation{Institute of Optics, University of Rochester, Rochester, NY 14627}

\author{Usman A. Javid}
\affiliation{Institute of Optics, University of Rochester, Rochester, NY 14627}

\author{Jingwei Ling}
\affiliation{Department of Electrical and Computer Engineering, University of Rochester, Rochester, NY 14627}

\author{Mingxiao Li}
\affiliation{Department of Electrical and Computer Engineering, University of Rochester, Rochester, NY 14627}

\author{Shixin Xue}
\affiliation{Department of Electrical and Computer Engineering, University of Rochester, Rochester, NY 14627}

\author{Kerry Vahala}
\affiliation{T.J. Watson Laboratory of Applied Physics, California Institute of Technology, Pasadena, California 91125, USA}

\author{Qiang Lin}
\email[Electronic mail: ]{qiang.lin@rochester.edu}
\affiliation{Department of Electrical and Computer Engineering, University of Rochester, Rochester, NY 14627}
\affiliation{Institute of Optics, University of Rochester, Rochester, NY 14627}

\begin{abstract}

Microwave signal generation with fast frequency tuning underlies many applications including sensing, imaging, ranging, time keeping, wireless communication, and high-speed electronics. Soliton microcombs are a promising new approach for photonic-based microwave signal synthesis.  To date, however, tuning rate has been limited in microcombs (and in frequency combs generally).  Here, we demonstrate the first microwave-rate soliton microcomb whose repetition rate can be tuned at a high speed. By integrating an electro-optic tuning/modulation element into a lithium niobate comb microresonator, a modulation bandwidth up to 75~MHz and a continuous frequency modulation rate up to $5.0\times 10^{14}$~Hz/s are achieved, several orders-of-magnitude faster than existing microcomb technology. These features are especially useful for disciplining an optical VCO to a long-term reference such as an optical clock or for tight phase lock in dual-microcomb clocks or synthesizers. Moreover, the microwave signal rate can be set to any X - W band rate. Besides its positive impact on microwave photonics, fast repetition rate control is generally important in all applications of frequency combs.  

\end{abstract}

\maketitle 

Several photonic technologies can produce coherent microwaves including optoelectronic oscillators \cite{maleki2011optoelectronic}, dual-frequency lasers \cite{pillet2008dual}, and Brillouin lasers \cite{li2013microwave}. However, the highest frequency stability microwave signals generated by any technology (electronic or optical) are derived from frequency comb technology using the principle of optical frequency division \cite{diddams2020optical}. Microcombs \cite{kippenberg2018dissipative} provide a powerful way to miniaturize this approach to the chip scale. And the need for detectable rate microcombs in comb systems has led to soliton microcombs with repetition rates in the radio and microwave frequency regimes \cite{herr2014temporal, liang2015high, yi2015soliton, suh2018gigahertz, liu2020photonic, lucas2020ultralow, yang2021dispersive, wang2021towards}. The continuous tuning of their repetition rate relies upon thermal or pump frequency control that is typically limited to audio bandwidths. Here we demonstrate a soliton microcomb whose microwave repetition rate can be continuously tuned with a record bandwidth of 75~MHz and a frequency modulation (FM) rate $> 5 \times 10^{14}$~Hz/s.
Besides application to microwave photonics, this high-speed rate control will be useful in all microcomb applications including frequency synthesizers \cite{spencer2018optical} and optical clocks \cite{Papp14,newman19}.    

\begin{figure*}[htbp]
	\centering\includegraphics[width=2\columnwidth]{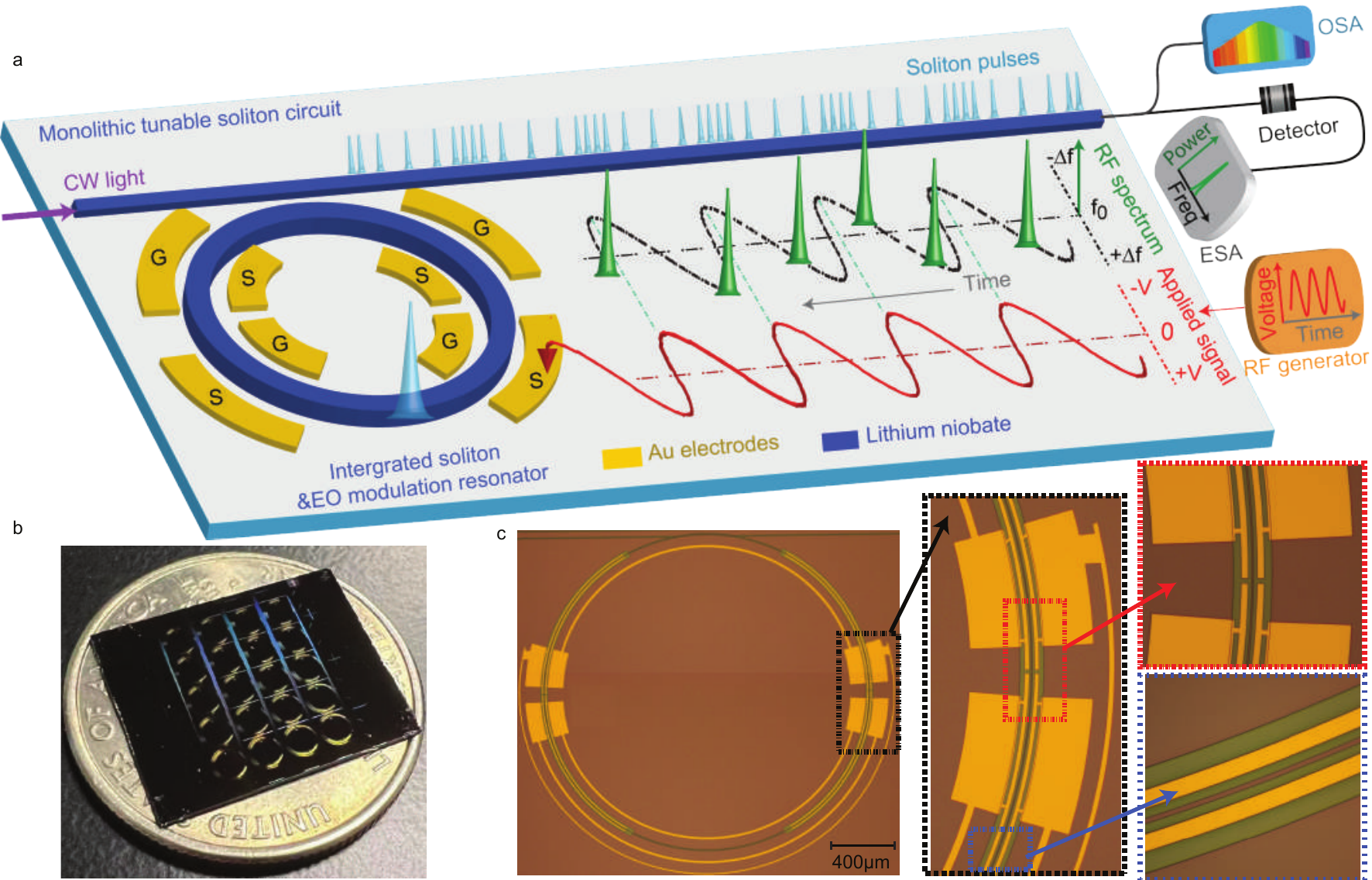}
	\caption{{\bf Concept of the high-speed tunable microwave-rate soliton source.}. {\bf (a)} Schematic of the tunable soliton source and its operational principle. CW: continuous-wave. OSA: optical spectrum analyzer. ESA: electrical spectrum analyzer. {\bf (b)} Photo of an LN comb resonator chip. {\bf (c)} Optical images of a device and the detailed structure of the driving electrodes and resonator waveguide. } 
	\label{Fig1}
	\end{figure*}

The device is an on-chip high-Q lithium niobate (LN) microresonator whose dispersion is engineered for soliton comb generation. The lithium niobate platform has recently been shown to enable self-starting and bi-directional switching of soliton states with the assistance of the photorefractive effect \cite{he2019self, gong2019soliton}. LN also exhibits a strong electro-optic Pockels effect, and here it is used for rapid tuning of the soliton repetition rate, by directly integrating electro-optic tuning and modulation elements into the comb resonator. Figure \ref{Fig1}(a) shows the device concept wherein electro-optic tuning and modulation elements are integrated into the comb resonator. In essence, the LN resonator functions simultaneously as a soliton comb generator and a high-speed electro-optic (EO) modulator. As shown below, this fully integrated device not only enables high-speed tuning of the soliton repetition rate, but also provides a way to tightly lock the repetition rate to an external microwave reference.

Figure \ref{Fig1}(b) shows the LN chip, which was fabricated on a z-cut LN-on-insulator (LNOI) wafer platform. Figure \ref{Fig1}(c) shows the detailed structure of a device which consists of a microring resonator, a pulley bus waveguide, and driving electrodes (Fig.~\ref{Fig1}(a)). The ring resonator is designed to have a waveguide width of 2.2~${\mu}$m, which yields a group velocity dispersion of about -0.035~${\rm ps^2/m}$ in the telecom band for the fundamental quasi-transverse-electric (quasi-TE) mode family that is suitable for soliton generation. The electrodes are placed along the ring resonator waveguide with an electrode-waveguide spacing of about 4~${\mu}$m so as to optimize the electro-optic tuning/modulation efficiency without impacting the optical Q of the resonator. The electrodes are 525~nm thick and 5~${\mu}$m wide in order to support high-speed modulation. The LN microring resonator exhibits an intrinsic optical Q of $\sim$ 4 million. A similar optical Q is obtained for resonators ranging in radius up to 1.5~mm.

\begin{figure*}[hbtp]
	\centering\includegraphics[width=2\columnwidth]{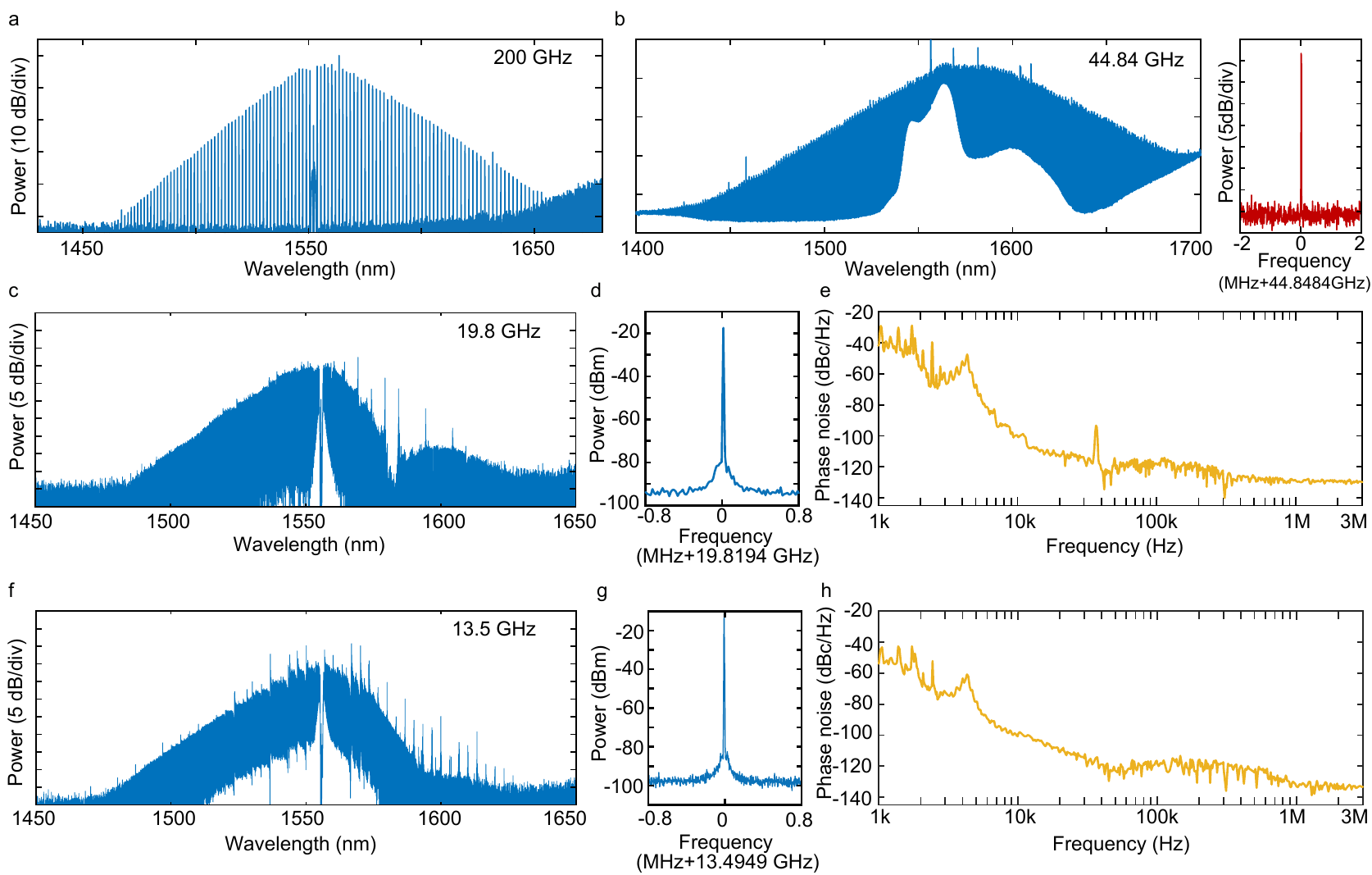}
	\caption{{\bf Soliton microcombs with different repetition rates}. {\bf (a), (b), (c), (f)} Optical spectra of soliton combs with repetition rates of 200, 44.84, 19.82, and 13.5~GHz, respectively, which are produced in LN comb resonators with radii of 100~${\mu}$m~{\bf (a)}, 450~${\mu}$m~{\bf (b)}, 1020~${\mu}$m~{\bf (c)}, and 1500~${\mu}$m~{\bf (f)}. The corresponding on-chip pump power is 33~mW, 396~mW, 282~mW, and 400~mW. The right panel in {\bf (b)} shows the electrical spectrum of the detected microwave signal from the 44.84~GHz soliton comb. The resolution bandwidth (RBW) of the RF spectrum is 500~Hz. {\bf (d) and (e)} Electrical spectrum and phase noise of the detected 19.82~GHz microwave signal produced by the soliton comb shown in {\bf (c)}. {\bf (g) and (h)} Same as {\bf (d) and (e)} but for the 13.5~GHz soliton comb shown in {\bf (f)}. In {\bf (d) and (g)}, the RBW of the RF spectrum is 200~Hz. In all figures, the devices are free running without active feedback, and the pump-laser-cavity detuning is self-stabilized by the photorefractive effect.} 
	\label{Fig2}
\end{figure*}


By increasing the radius of the ring resonator from 100~${\mu}$m to 450~${\mu}$m, we are able to vary the soliton repetition rate $f_{r}$ from 200~GHz to 44.84~GHz, as shown in Fig.~\ref{Fig2}(a) and (b). Further increase of resonator size is non trivial due to the interference of stimulated Raman scattering. LN exhibits rich Raman scattering characteristics that were recently shown to introduce self-frequency shift (SFS) on the Kerr solitons  \cite{he2019self}. However, Raman lasing is also possible when a Stokes frequency matches a cavity resonance \cite{gong2019soliton, yu2020raman}. For a resonator with a radius $\ge$1~mm, the free-spectral range is small enough ($\le$20~GHz) that Raman lasing becomes unavoidable, and this perturbs soliton generation. For the z-cut resonators used here, the most significant interference comes from the Stokes wave with a Raman frequency shift of $\sim$19~THz (${\rm A_1 (TO_4)}$ phonon mode of LN \cite{basiev1999raman, sanna2015raman}. To resolve this issue, the pulley bus waveguide is designed to be critical coupled at the pump wavelength around 1550~nm and over coupled at the Raman Stokes wavelength around 1720~nm (Supplementary Information (SI)). As a result, Raman lasing can be suppressed. 

Figure \ref{Fig2}(c) and (f) show two examples of soliton microcombs with $f_r$ of 19.82~GHz and 13.5~GHz, respectively. The slight spectral distortion around 1580~nm is a side effect of the bus waveguide which is designed to be under coupled at this wavelength (see SI). This side effect, however, does not impact the integrity of the Kerr solitons and their coherence is evident in the detected microwave signal shown in Fig.~\ref{Fig2}(d) and (g). Here, both microwave signals at 19.82 and 13.5~GHz exhibit a signal-to-noise ratio greater than 70~dB. As shown in Fig.~\ref{Fig2}(e) and (h), the phase noise of the microwaves is about -40~dBc/Hz at 1~kHz, -110~dBc/Hz at 10~kHz, and below -130~dBc/Hz at 3~MHz. These phase noise levels are comparable to those demonstrated recently in other on-chip soliton platforms \cite{liu2020photonic, yang2021dispersive}. The spectral bump around 4~kHz is likely due to the impact of the frequency noise of the pump laser (New Focus, TLB-6328) \cite{liu2020photonic, yang2021dispersive}.

\begin{figure*}[htpb]
	\centering\includegraphics[width=2\columnwidth]{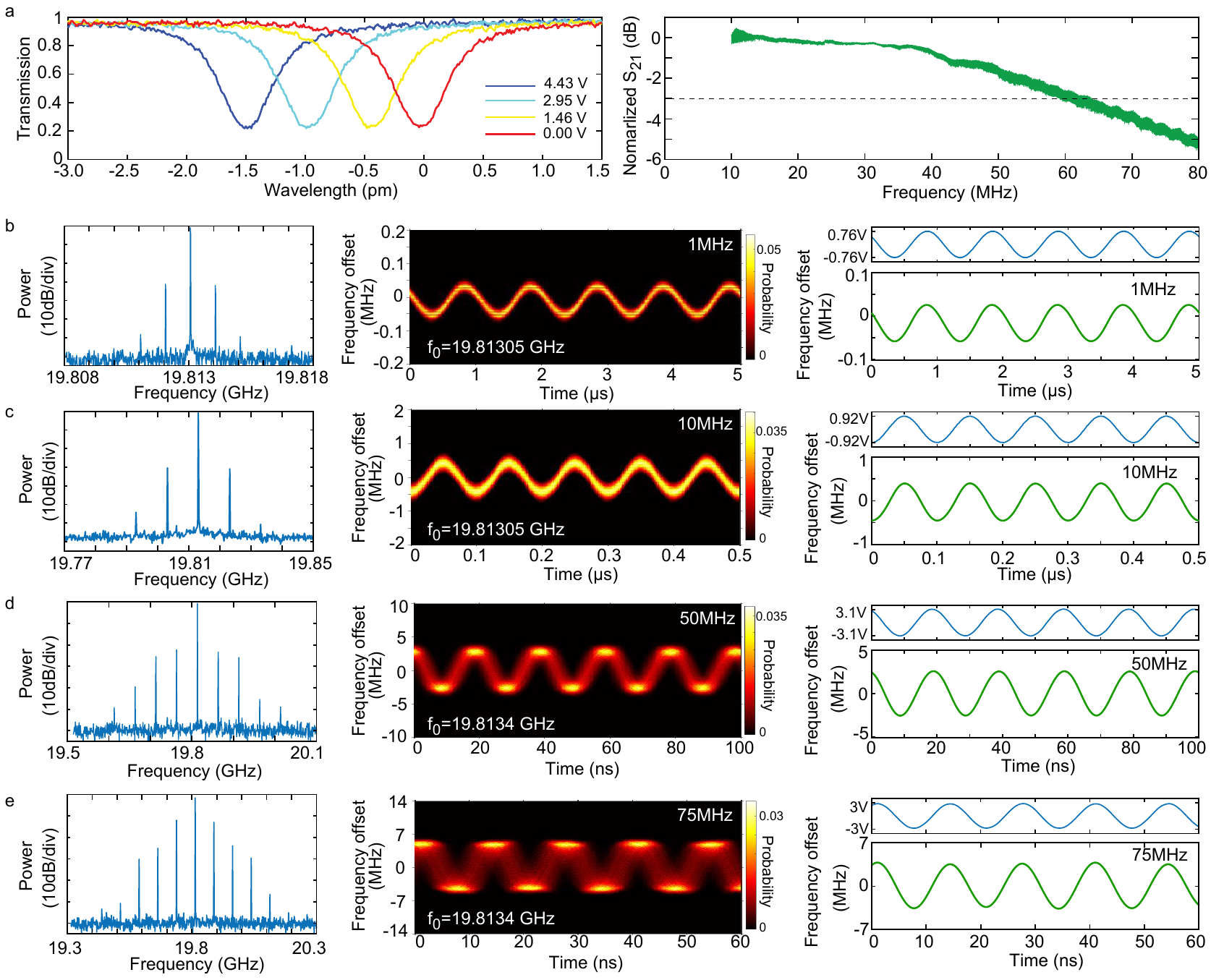}
	\caption{{\bf High-speed modulation of the soliton microcombs. } {\bf (a) } Electro-optic tuning performance of the comb resonator. Left: DC tuning of a cavity resonance. Right: Electro-optic modulation response ${\rm S_{21}}$ of the comb resonator. To characterize the EO tuning performance, the pump laser power was reduced to a low level such that no comb or other nonlinear effect was produced, and the device functioned essentially as a pure electro-optic modulator. To obtain the ${\rm S_{21}}$ curve, the pump laser wavelength is fixed to a cavity resonance, and the modulation amplitude of resonator transmission was recorded as a function of the modulation frequency with a network analyzer. {\bf (b),(c),(d),(e)} Frequency modulation of the soliton comb repetition rate at a modulation frequency of 1 {\bf (b)}, 10 {\bf (c)}, 50 {\bf (d)}, and 75~MHz {\bf (e)}, respectively. Left column: spectrum of the detected microwave. Center column: frequency vs time spectrum. Right column: top: applied driving voltage; bottom: time-dependent frequency curve, which is the averaged trace of the frequency vs time spectrum shown in the center column. The data were recorded with an electrical spectrum analyzer (Tektronics, RSA5126B). In the frequency vs time spectra shown in the center column, the blurring of the spectrum with increased modulation frequency is due to the limited bandwidth (160~MHz) of the spectrum analyzer. In all figures, same as Fig.~\ref{Fig2}, the device is free running without active feedback, and the pump-laser-cavity detuning is self-stabilized by the photorefractive effect. } \label{Fig3}
\end{figure*}

\begin{figure*}[htbp]
	\centering\includegraphics[width=2\columnwidth]{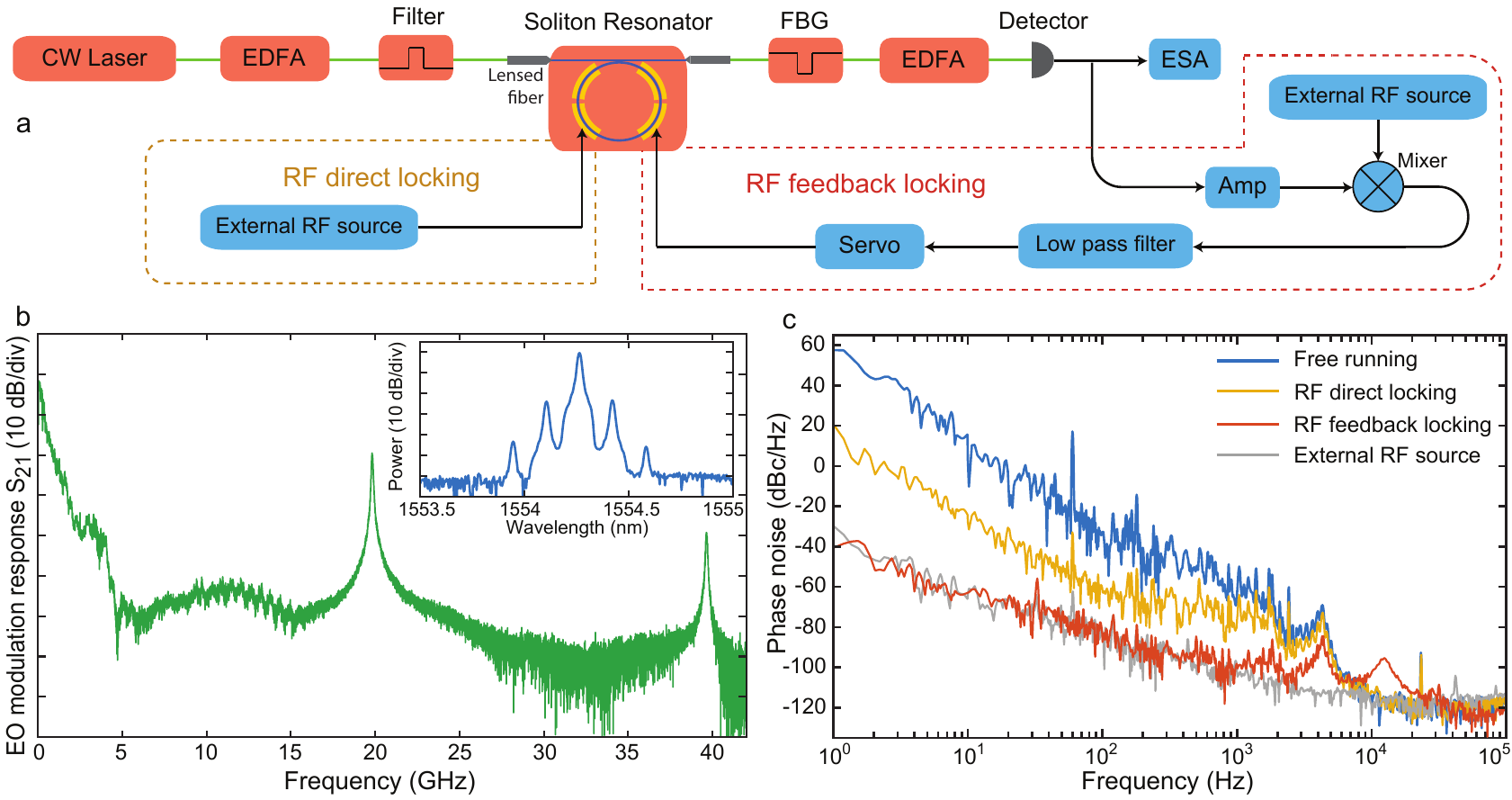}
\caption{{\bf Locking of soliton repetition rate to a reference microwave source.} {\bf (a)} Schematic of the experimental testing setup. EDFA: erbium-doped fiber amplifier. FBG: fiber Bragg grating filter. Amp: electrical amplifier. The soliton repetition rate is locked to an external reference microwave source (Anritsu, MG3697C) via two separate approaches. In the first approach, the external microwave signal directly drives the comb resonator at a modulation rate close to the resonator FSR (yellow dashed box). This is labeled as RF direct locking. In the second approach, the detected soliton microwave signal is compared with the external reference microwave and the error signal is used to electro-optically tune the soliton repetition rate (red dashed box). This is labeled as RF feedback locking. {\bf (b)} Electro-optic modulation response $S_{21}$ of the comb resonator over a broad frequency range. The inset shows the optical spectrum of the produced comb-like sidebands when the comb resonator was driven with a 19.81~GHz microwave signal with a power of 20~dBm. The $S_{21}$ curve was recorded at a low laser power such that the device functions as a pure electro-optic resonator, similar to Fig.~\ref{Fig3}(a). {\bf (c)} Phase noise spectrum of the detected 19.81~GHz microwave. The yellow and red curves show the case of RF direct locking (yellow box in {\bf (a)}) and RF feedback locking (red box in {\bf (a)}), respectively, while the blue curve shows the case when the device is free running. The gray curve shows the phase noise of the external reference microwave. }\label{Fig4}
\end{figure*}

In order to dynamically tune and modulate the microwave-rate solitons, we integrate EO tuning/modulating components directly onto the comb resonator as shown in Fig.~\ref{Fig1}(a). For the z-cut resonator, we utilize the $r_{22}$ electro-optic tensor element of LN to tune the fundamental quasi-TE modes. For a circularly shaped ring resonator, our analysis shows that the EO tuning efficiency can be maximized with three groups of driving electrodes, each of which contains two pairs of signal-ground electrodes and spans an angle of 60 degrees (see SI for details). A detailed device layout is shown in Fig.~\ref{Fig1}(c), in which only two groups of electrodes are fabricated so as to avoid the interference with the coupling bus waveguide (see also Fig.~\ref{Fig1}(a)). To simplify experimental testing, we utilized only one group of electrodes for EO tuning and modulation. Nonetheless, the integrated EO tuning still achieves good cavity resonance tuning efficiency of 0.34~pm/V, as shown in Fig.~\ref{Fig3}(a). The same figure also shows the EO-modulation response of the comb device, which gives a 3~dB bandwidth of about 61~MHz, corresponding to the photon lifetime limit of the comb resonator.

The broadband EO response of the device implies that high-speed tuning control of the soliton microcomb is possible. To show this feature, we applied a sinusoidal electric signal to the 19.81~GHz comb resonator and monitored the frequency of the detected microwave signal. As shown in Fig.~\ref{Fig3}(b)-(e), the sinusoidal EO-drive produces a sinusoidal frequency modulation (FM) of the microwave signal. At a modulation frequency of 1~MHz (Fig.~\ref{Fig3}(b)), a peak driving voltage of ${\rm V_p}$=0.76V produces an FM amplitude of 41.8~kHz which corresponds to a FM efficiency of 55~kHz/V. The FM efficiency increases considerably with the modulation frequency, reaching a value of 463~kHz/V and 824~kHz/V at the modulation frequency of 10 and 50~MHz, respectively (Fig.~\ref{Fig3}(c) and (d)). As shown in Fig.~\ref{Fig3}(e), we are able to modulate the microwave signal at a frequency as high as 75~MHz, where a driving voltage of ${\rm V_p}$ = 3.0~V produces a FM amplitude of 3.45~MHz, corresponding to an FM efficiency of 1.15~MHz/V. Here, the blurring of the time-frequency spectrum is due to the limited bandwidth (160~MHz) of the spectrum analyzer (Tektronics, RSA5126B) used for the time-dependent frequency characterization. For the same reason, the time-dependent frequency analysis likely underestimates the FM amplitude since it only captures the first-order modulation sidebands at such a high modulation frequency. Indeed, the microwave spectrum shown in Fig.~\ref{Fig3}(e) implies a considerably higher FM amplitude and efficiency, whose details are provided in the SI. As such an FM amplitude is realized within a time scale of only $\sim$6.7~ns (half of the modulation period), the frequency modulation rate reaches a value of $> 5\times 10^{14}$~Hz/s at the modulation frequency of 75~MHz. 
The FM efficiency drops, however, with further increases in modulation frequency, due to the photon lifetime limit of the resonator. Further details are provided in the SI. 
 
One mechanism responsible for the observed FM of the microwave signal is the Raman-induced SFS of the solitons whose magnitude depends on the laser-cavity detuning \cite{yi2016theory}. EO modulation of the comb resonator modulates the laser-cavity detuning of the pump wave which in turn changes the magnitude of SFS and thus shifts the carrier frequency of the Kerr solitons. Due to the group-velocity dispersion of the resonator, such a shift of soliton carrier frequency translates into a change of the repetition rate. As shown in the SI, this mechanism accounts for an FM efficiency of $\sim$(100--200)~kHz/V, which explains well the observed phenomeona at low modulation frequencies. However, the FM efficiencies observed at higher modulation frequencies of 50 and 75~MHz are considerably larger than this value. The underlying reason is likely related to the speed of EO modulation which becomes comparable to the photon lifetime in the resonator so that the cavity resonances cannot adiabatically follow the EO modulation anymore. However, the exact physical mechanism is not clear at this moment and will require further exploration. On the other hand, the same mechanism of Raman-induced SFS leads to a certain extent of amplitude modulation of the produced microwave, whose details are provided in the SI.

The actual EO response of the device is much larger than the cavity bandwidth as evidenced in Fig.~\ref{Fig4}(b).  In this measurement, the resonantly enhanced EO response is apparent at the modulation frequencies of 19.81 and 39.62~GHz corresponding to one and two free-spectral ranges of the cavities. Comb-like sidebands \cite{zhang2019broadband} can be produced by driving the comb resonator at a frequency of 19.81~GHz (Fig.~\ref{Fig4}(b), inset). Apparently, the device offers an EO bandwidth up to tens of gigahertz. With this broadband modulation response of the comb resonator, we are able to lock the soliton repetition rate in two ways. On one hand, we can apply the 19.81~GHz reference microwave directly to the comb resonator during the soliton generation (Fig.~\ref{Fig4}(a), yellow box). The produced EO comb then seeds the soliton generation. This approach is similar to the injection locking approach \cite{liu2020photonic}, but the reference microwave is now fed directly to the comb resonator itself rather than through external modulation on the pump laser. As shown in the yellow curve in Fig.~\ref{Fig4}(c), such a direct locking approach is able to suppress the phase noise by about 40~dB over the frequency range of 1~Hz -- 1~kHz. On the other hand, we can also compare the detected microwave with the reference oscillator and apply the error signal to electro-optically lock the repetition rate (Fig.~\ref{Fig4}(a), red box). As shown in the red curve of Fig.~\ref{Fig4}(c), this feedback locking approach is able to suppress the phase noise down to that of the reference microwave over the frequency range 1~Hz -- 3~kHz. The residual peaks around 10~kHz are due to the bandwidth limit of the servo unit.  

The current comb resonator chip is not packaged and uses lensed fiber to couple the pump laser onto the chip (Fig.~\ref{Fig4}(a)). This can induce fluctuations in the coupled pump power to the comb resonator (especially at low frequencies). These fluctuations could be substantially reduced in the future with packaging of the chip \cite{liu2020photonic, yang2021dispersive}. On the other hand, the photorefractive effect of LN could have a potential impact on the phase noise at low frequencies, whose exact behavior is difficult to characterize at this moment and will require further exploration in the future. 


\section*{Discussion}


It was shown recently \cite{yi2017single} that soliton microcombs exhibit a certain "quiet point" around which the phase noise of the microwave can be significantly suppressed due to the recoil effect between soliton SFS and dispersive wave produced via mode crossing. This approach was not implemented in our current experiments which focus on the basic phase-noise performance offered by a LN soliton microcomb and the high-speed tunability of soliton repetition rate. Use of the quite-point method could further improve the phase-noise performance of the LN soliton microcombs.  

In summary, we have demonstrated the first microwave-rate soliton microcomb whose repetition rate can be tuned at a high speed. By taking advantage of the strong electro-optic Pockels effect of LN and by integrating electro-optic tuning and modulation components directly into the LN comb resonator, we are able to achieve a continuous frequency modulation speed up to 75~MHz and a frequency modulation rate up to $> 5\times 10^{14}$~Hz/s. The device exhibits a modulation efficiency of $>1$~MHz/V, which could be further increased if all three groups of driving electrodes are employed. Using the fast rate control, servo locking of the comb rate to an external microwave reference was demonstrated. Also, injection locking of the comb rate by electro-optical modulation using a microwave reference was demonstrated. The demonstrated device brings high speed modulation to soliton microcombs, providing a new approach to electro-optic processing of coherent microwaves. 

\section*{Methods}
\subsection{Device fabrication}
The devices were fabricated on a 610-nm z-cut LN-on-insulator (LNOI) wafer. Ring resonators and waveguides structures were defined by electron-beam lithography with ZEP520a as resist, followed by etching to about 410~nm depth with Ar ion milling. After removing the ZEP520a residue, 525~nm thick Au electrodes were patterned by electron-beam lithography with PMMA as resist, followed by deposition using an electron beam evaporator. An overnight lift-off process was applied to remove PMMA and residual Au.

\textbf{Data availability.} The data that support this study are available from the authors on reasonable request.

\section*{Acknowledgments}

We thank Qingxi Ji at Caltech for helpful discussions on RF feedback locking experiments. We also thank Prof.~Lin Chang at Peking University for helpful discussions. This work is supported in part by the Defense Threat Reduction Agency-Joint Science and Technology Office for Chemical and Biological Defense (grant No. HDTRA11810047), the Defense Advanced Research Projects Agency (DARPA) LUMOS program under Agreement No. HR001-20-2-0044, and the National Science Foundation (NSF) (ECCS-1810169, ECCS-1842691
and, OMA-2138174). This work was performed in part at the Cornell NanoScale Facility, a member of the National Nanotechnology Coordinated Infrastructure (National Science
Foundation, ECCS-1542081); and at the Cornell Center for Materials Research (National Science Foundation, Grant No.~DMR-1719875). 

The project or effort depicted was or is sponsored by the Department of the Defense, Defense Threat Reduction Agency. The content of the information does not necessarily reflect the position or the policy of the federal government, and no official endorsement should be inferred.

\section*{Author contributions}
Y.H. designed and fabricated the sample. Y.H. designed and performed the experiments. Y.H., R.L., and U.J. did the numerical simulations. Y.H. and R.L. analyzed the data. U.J.,J.L., R.L., and M.L. assisted in the experiments. X.X. assisted in device fabrication. Y.H., R.L., K.V., and Q.L. wrote the manuscript. K.V. and Q.L. supervised the project. Q.L. conceived the device concept. 

\section*{Additional information}
Correspondence and requests for materials should be addressed to Q.L.

\textbf{Competing financial interests:} The authors declare no competing financial interests.

\bibliographystyle{naturemag}
\bibliography{References}

\clearpage

\setcounter{section}{0}
\setcounter{subsection}{0}
\section*{Appendix}

\subsection{Design of the pulley coupling waveguide}

\begin{figure}[htpb]
	\centering\includegraphics[width=1\linewidth]{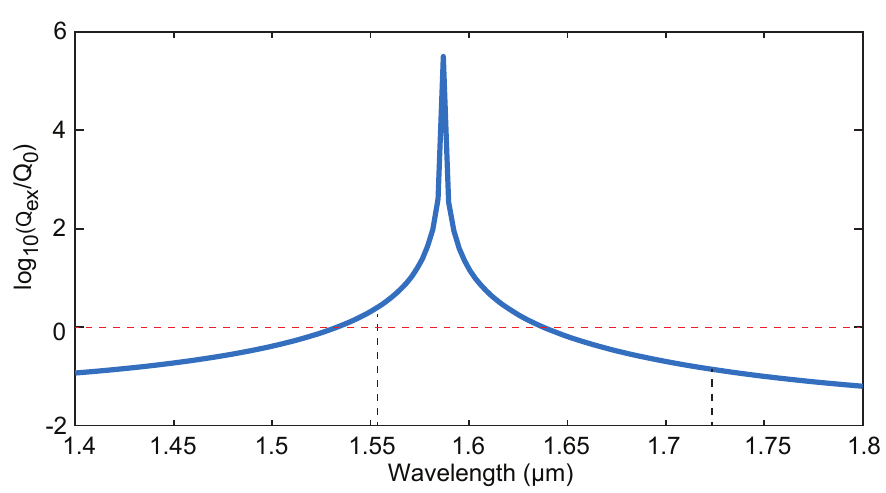}
	\caption{{Simulated ratio of external coupling quality factor $Q_{ex}$ to intrinsic quality factor $Q_0$} for the designed pulley bus waveguide. } \label{Fig5}
\end{figure}

To prevent the Raman lasing, we design the pulley coupling waveguide such that the resonator is close to critical coupling around the pump wavelength of 1550nm but it is strongly over-coupled at the Raman Stokes wavelength around 1720~nm. The effective refractive index of the waveguides is modeled by the finite-element method via COMSOL, and the coupling condition of the bus waveguide can be simulated with a coupled-mode theory \cite{moille2019broadband}. Our detailed modelings show that the desired coupling condition can be obtained with a bus-waveguide width of 1.765~${\mu}$m, a pulley angle of 10 degrees, and a constant gap of 300~nm between the bus waveguide and the ring resonator in the pulley coupling region. Figure \ref{Fig5} shows the simulated ratio of the external coupling Q, $Q_{ex}$, to the intrinsic optical Q, $Q_0$. It shows that such a pulley waveguide design is able to achieve nearly critical coupling at the pump wavelength of 1550~nm but is strongly over coupled at the Raman Stokes wavelength around 1720~nm. A side effect of such a design is that a coupling resonance appears around a wavelength of 1580~nm around which the resonator is deeply under coupled. This side effect is responsible for the spectral distortion of the soliton combs shown in Fig.~\ref{Fig2}(c) and (f) of the main text. The side effect can be removed by further optimizing the pulley waveguide design. A similar approach was used in Ref.~\cite{gong2019raman}.

\subsection{Design of electro-optic tuning/modulation element for the soliton comb resonator}

\noindent{\bf A general theory of the electro-optic effect in an  LN electro-optic resonator:} Here we provide first a general theory describing the electro-optical (EO) effect in an LN electro-optic resonator, and then apply it for designing the EO tuning/modulation elements for the z-cut soliton comb resonator. 

In general, a driving electric field $\ket{\mathcal{E}}$ applied to the LN medium will perturb the relative permittivity tensor $\olr{\epsilon_r} = (\epsilon_r)_{ij}$ and introduce a change $\delta\olr{\epsilon_r} = (\delta\epsilon_r)_{ij}$,
\begin{equation}
(\delta\epsilon_r)_{ij} = - (\epsilon_r)_{ik} r_{klm} \scE_m (\epsilon_r)_{lj}.
\label{eq:delep}
\end{equation}
To obtain Eq.~(\ref{eq:delep}), we have used the electro-optic Pockels tensor $r_{ijk} \equiv (\partial \eta_{ij}/\partial \scE_k)_{\scE=0}$ where $\eta_{ij}$ is the impermeability tensor which is related to the relative permittivity tensor $(\epsilon_r)_{ij}$ as $\eta_{ij} \equiv (\epsilon_r)_{ij}^{-1}$ \cite{yariv1984optical}. In the crystallographic coordinate system of LN, the relative permittivity tensor is given by
\begin{equation}
(\epsilon_r)_{ij} = 
\begin{pmatrix}
\epsilon _{11} & 0 & 0 \\
0 & \epsilon _{11} & 0 \\
0 & 0 & \epsilon _{33} \\
\end{pmatrix}, \label{eq:epsilon}
\end{equation}
where $\epsilon_{11}$ and $\epsilon_{33}$ are the relative permittivity coefficients in Cartesian coordinates, and the $z$-axis corresponds to the optic axis of LN. Accordingly, the Pockels tensor $r_{ijk}$ has a contracted form $r_{ij}$ \cite{yariv1984optical}:
\begin{equation}
	r_{ij} = 
\begin{pmatrix}
	0 & -r_{22} & r_{13} \\
	0 & r_{22} & r_{13} \\
	0 & 0 & r_{33} \\
	0 & r_{51} & 0 \\
	r_{51} & 0 & 0 \\
	-r_{22} & 0 & 0 
\end{pmatrix}. \label{eq:r_ik}
\end{equation}

The EO induced perturbation to the dielectric tensor would introduce a shift to the cavity resonance $\omega_0$, which can be obtained by the perturbation theory given as \cite{joannopoulos2011photonic, johnson2002perturbation} 
\begin{equation}
\delta\omega_0 = -\frac{\omega_0}{2} \frac{ \left< E \right| \delta\olr{\epsilon_r} \left| E \right>}{ \left< E \right|  \olr{\epsilon_r} \left| E \right>} = 
- \frac{\omega_0}{2} \frac{ \left< E_i, (\delta\epsilon_r)_{ij} E_j \right>}{ \left< E_i  , (\epsilon_r)_{ij} E_j \right>},
\label{eq:deltaom0}
\end{equation}
where $\ket{E}$ is the optical field vector of the cavity resonance mode. By using Eqs.~(\ref{eq:epsilon}) and (\ref{eq:r_ik}) in Eq.~(\ref{eq:delep}) and then substituting it into Eq.~(\ref{eq:deltaom0}), we obtain the resulting EO-induced cavity resonance shift as 
\begin{equation}
\delta\omega_0 =  - \frac{\omega_0}{2} \frac{\int_{\text{core}} \Diff3\mathbf{r} \sum_{u,v}{\xi_u^{(v)}}}{\int_{\text{all}}  \Diff3\mathbf{r} \bar{\xi} }, \quad u \in \{x, y, z\},~ v \in \{1, 2\},
\label{eq:dnu}
\end{equation}
where $\xi_u^{(v)}$ represents the fractional contribution from different polarization components given as:
\begin{equation}
\begin{split}
\xi_x^{(1)} &= 2 \mathcal{E}_x r_{51} \epsilon _{33} \epsilon _{11} \operatorname{Re}\left\{E_x E_z^*\right\}, \\
\xi_x^{(2)} &= -2 \mathcal{E}_x r_{22} \epsilon _{11}^2 \operatorname{Re}\left\{E_x E_y^*\right\}, \\
\xi_y^{(1)} &= \mathcal{E}_y r_{22} \epsilon _{11}^2 \left(\left| E_y\right|^2-\left| E_x\right|^2\right), \\
\xi_y^{(2)} &= 2 \mathcal{E}_y r_{51} \epsilon _{33} \epsilon _{11} \operatorname{Re}\left\{E_y E_z^*\right\}, \\
\xi_z^{(1)} &= \mathcal{E}_z  r_{13} \epsilon _{11}^2 \left(\left| E_x\right|^2+\left| E_y\right|^2\right), \\
\xi_z^{(2)} &= \mathcal{E}_z r_{33} \epsilon _{33}^2 \left| E_z\right|^2,
\end{split} \label{eq:eta}
\end{equation}
where $E_x$, $E_y$, and $E_z$ are the three polarization components of the optical field vector. $\bar{\xi}$ is the normalization factor for the optical field given as
\begin{equation}
\bar{\xi} = \epsilon _{11} \left(\left| E_x\right|^2+\left| E_y\right|^2\right)+\epsilon _{33} \left| E_z\right|^2, \label{eq:etabar}
\end{equation}
that is related to the energy density of the optical mode. In Eq.~(\ref{eq:dnu}), $\int_{\rm core}$ stands for the spatial integration over the LN waveguide layer only, and $\int_{\rm all}$ stands for that over the whole space. In $\xi_u^{(v)}$ (Eq.~(\ref{eq:eta})), the subscript $u$ ($u=x,y,z$) denotes the contribution from the $u$ polarization component of the driving electric field and the superscript index $v$ ($v=1,2$) denotes the contribution from different polarization components of the optcal cavity mode.

Equations (\ref{eq:dnu})-(\ref{eq:etabar}) can be used to described the electro-optic effect in any arbitrary LN electro-optic micro/nanoresonator, including microring, racetrack, microdisk, and photonic crystals. 

\noindent{\bf Design of EO tuning/modulation elements for the z-cut soliton comb resonator:} We now use the theory developed above to describe the electro-optic effect in a z-cut microring resonator. Specifically, we are interested in the EO effect on the fundamental quasi-TE modes of a circularly shaped z-cut microring resonator that we employ for producing soliton microcomb. 

\begin{figure}[b!]
	\centering
	\includegraphics[width=1\linewidth]{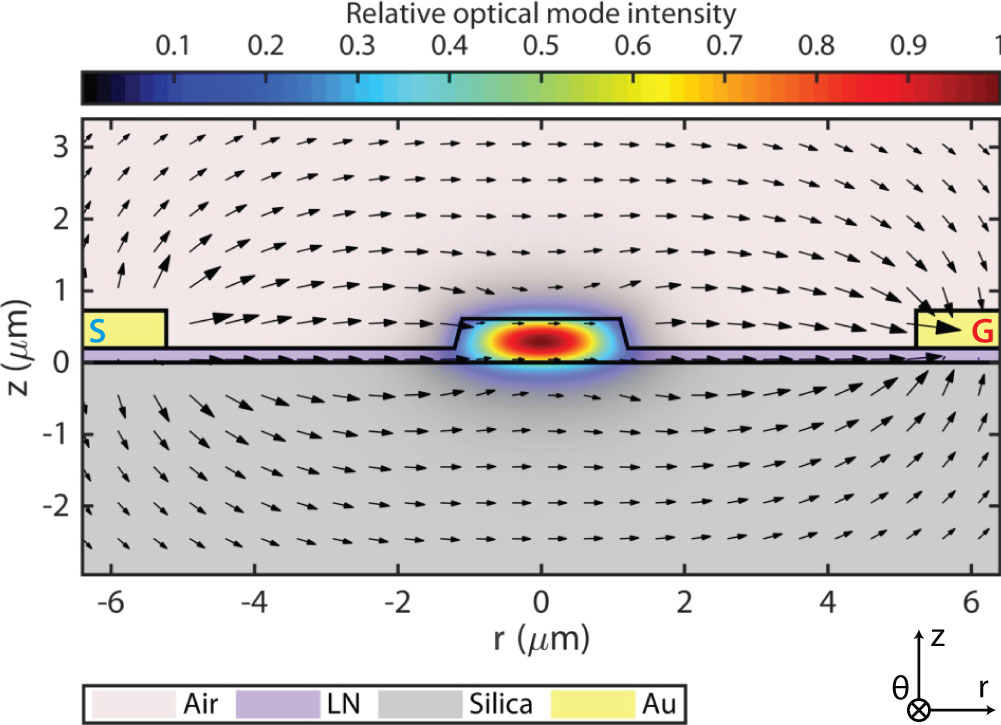}
	\caption[]{Optical mode field profile (color map) of the fundamental quasi-TE mode and vector map of applied electric field (arrow map) of the LN microring comb resonator, simulated by the finite-element method. }
	\label{fig:comsol-mode}
\end{figure}

Figure~\ref{fig:comsol-mode} shows the optical mode field profile of a fundamental quasi-TE mode in such a resonator. It shows clearly that the cavity mode has its polarization dominantly lying in the device plane (the $x$-$y$ plane) along the radial direction, with $E_z$ component negligible. The optical field $\ket{E}$ can thus be approximated as:
\begin{equation}
\ket{E} \approx E_0(r,z) \left( \cos\theta \hat{x} + \sin\theta \hat{y} \right) e^{i m \theta},
\label{eq:eopt}
\end{equation}
where $E_0(r,z)$ is the optical field amplitude, $m$ is the mode number of the cavity mode, and $\hat{x}$  $(\hat{y})$ is the unit vector in the $x$ ($y$) axis. Due to the rotation symmetry of device, we have adopted a cylindrical coordinate system ($r,\theta,z$) in Eq.~(\ref{eq:eopt}).

\begin{figure}[b!]
	\centering
	\includegraphics[width=0.8\linewidth]{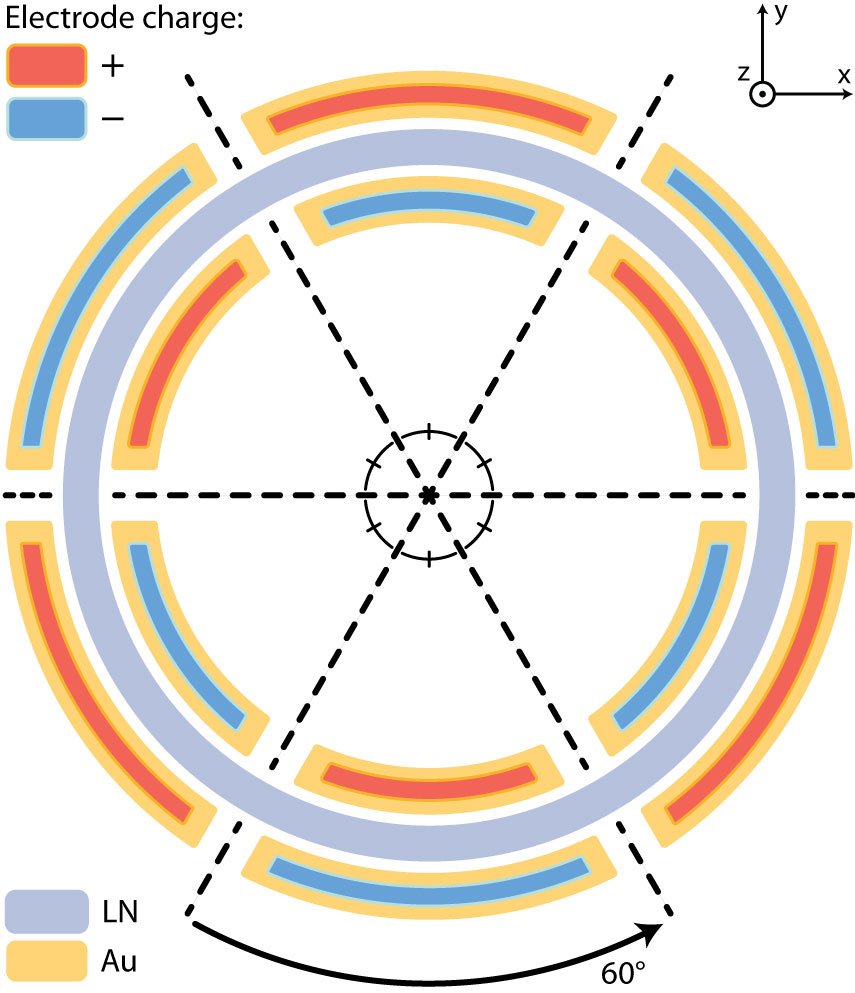}
	\caption[]{Layout of the driving electrodes for a $z$-cut microring resonator. There are six pairs of electrodes with alternating polarities all arranged symmetrically and spaced azimuthally by $60^o$ along the ring circumference. Colors indicate Lithium niobate (LN), gold (Au), or electrode charge ($+$ or $-$).}
	\label{fig:zcut}
\end{figure}

For the EO tuning and modulation, we focus on the scenario that the electrodes are placed on the opposite sides of the ring waveguide (Fig.~\ref{fig:comsol-mode}), which is easy to implement in practice. For this case, as shown in Fig.~\ref{fig:comsol-mode}, the driving electric field inside the LN waveguide core also dominantly lies in the $x$-$y$ plane and along the radial direction. As such, the driving electric field vector can be approximated as 
\begin{equation}
\ket{\scE} \approx \scE_0 (r,z) \left( \cos\theta \hat{x} + \sin\theta \hat{y} \right),
\label{eq:eac}
\end{equation}
where $\scE_0(r,z)$ is the driving field amplitude.

\begin{figure}[hbp!]
	\centering
	\includegraphics[width=1\linewidth]{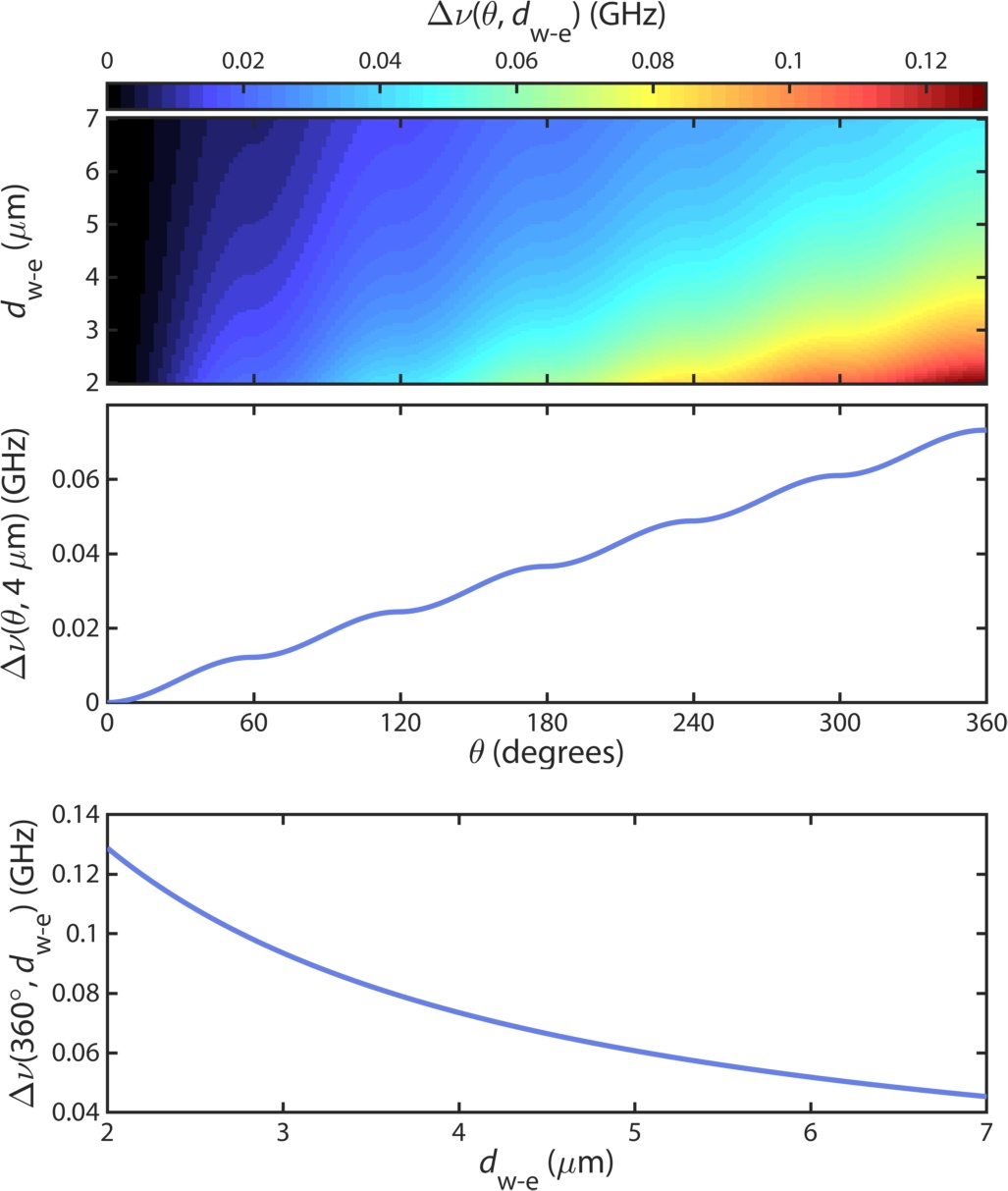}
	\caption[]{Simulated frequency shift $\Delta\nu$ vs.\ azimuthal angle $\theta$ and waveguide-electrode spacing $d_{\text{w-e}}$ with an applied voltage of 1~V, for a resonance frequency of $\sim$ 193.4 THz (1.55 $\mu m$). Top: surface plot vs.\ $\theta$ and $d_{\text{w-e}}$ where the colormap represents the frequency shift $\Delta\nu$. Middle: $\Delta\nu$ vs.\ electrode coverage angle $\theta$ for $d_{\text{w-e}} \sim 4$ $\mu m$. Bottom: $\Delta\nu$ vs.\ waveguide-electrode spacing $d_{\text{w-e}}$ with a full $2\pi$ microring electrode coverage.}
	\label{fig:eom-sim}
\end{figure}

With the optical field and driving electric field given in Eqs.~(\ref{eq:eopt}) and (\ref{eq:eac}), Eq.~(\ref{eq:eta}) shows that, among all the terms, only $\xi_x^{(2)}$ and $\xi_y^{(1)}$ are non-negligible, which are given by
\begin{eqnarray}
\xi_x^{(2)} &=& - r_{22} \epsilon _{11}^2 \scE_0 (r,z) |E_0(r,z)|^2 \sin(2\theta) \cos \theta, \label{eta_x2_approx} \\
\xi_y^{(1)} &=& - r_{22} \epsilon _{11}^2 \scE_0 (r,z) |E_0(r,z)|^2 \sin\theta \cos(2 \theta). \label{eta_y1_approx}
\end{eqnarray}
As a result, the EO-induced cavity resonance shift (Eq.~(\ref{eq:dnu})) becomes
\begin{eqnarray}
\delta\omega_0 &=&  \frac{\omega_0}{4\pi} \frac{\int_{\rm core}{r_{22} \epsilon_{11}^2 \scE_0 |E_0|^2 rdrdz} \int_{\theta_1}^{\theta_2} {\sin(3\theta) d\theta}}{\int_{\rm all} {\epsilon_{11} |E_0|^2 rdrdz} } \nonumber \\
&=&  \frac{\omega_0}{4\pi} \frac{\int_{\rm core}{r_{22} \epsilon_{11}^2 \scE_0 |E_0|^2 rdrdz} }{\int_{\rm all} {\epsilon_{11} |E_0|^2 rdrdz} } \frac{1}{3}\left[\cos(3\theta_1) -\cos(3\theta_2) \right], \nonumber \\
\label{eq:dnu_approx}
\end{eqnarray}
where we have assumed that the driving electrodes span over an azimuthal angle range between $\theta_1$ and $\theta_2$. Apparently, a driving electrode spans over the entire circular ring would results in a zero net frequency shift. However, Eq.~(\ref{eq:dnu_approx}) shows that $\delta \omega_0$ exhibits a period of $120^o$ and it reaches a peak value by choosing $(\theta_1, \theta_2) = (0, 60^o)$. Moreover, $\delta \omega_0$ remains at the same peak value for $(\theta_1, \theta_2)$ = $(0,60^o)$, $(120^o,180^o)$, or $(240^o,300^o)$ but flips its sign for $(\theta_1, \theta_2)$ = $(60,120^o)$, $(180^o,240^o)$, or $(300,360^o)$. Therefore, we can maximize the magnitude of the EO-induced resonance frequency shift by alternating the sign or polarity of the driving field across six identical electrode pairs placed consecutively along the microring circumference. Figure \ref{fig:zcut} shows the arrangement of the driving electrodes. With this approach, the induced frequency shifts by the six electrode sections all add up constructively, resulting in a maximal EO frequency shift of 
\begin{eqnarray}
\delta\omega_0 =  \frac{\omega_0}{\pi} \frac{\int_{\rm core}{r_{22} \epsilon_{11}^2 \scE_0 |E_0|^2 rdrdz} }{\int_{\rm all} {\epsilon_{11} |E_0|^2 rdrdz} }. \label{eq:dnu_approx_max}
\end{eqnarray}

To verify the function of the proposed EO modulation structure shown in Fig.~\ref{fig:zcut} and to quantify the magnitude of resulting EO tuning, we used the full vectorial form of the optical field and that of the driving electric field simulated by the finite-element method, in Eqs.~(\ref{eq:dnu})-(\ref{eq:etabar}) to find the induced resonance frequency shift. The results are presented in Fig.~\ref{fig:eom-sim}, which show the dependence of $\Delta\nu = \delta \omega_0/(2\pi)$ on both the electrode coverage angle $\theta$ and the waveguide-electrode spacing $d_{\text{w-e}}$, with an applied voltage of 1~V. Figure \ref{fig:eom-sim}(b) shows that, for a waveguide-electrode spacing of 4~${\rm \mu m}$, each pair of signal-ground driving electrodes offers an frequency tuning efficiency of $\sim$12~MHz/V. As a result, a group of two-pairs of driving electrodes, as we did in our device, will produce a frequency tuning efficiency of $\sim$24MHz/V, corresponding to a wavelength tuning efficiency of $\sim$0.19~pm/V which is close to what we measured on the fabricated device. In particular, the driving electrode structure proposed in Fig.~\ref{fig:zcut} does offer constructive EO tuning effect from the six pairs of the electrodes, with $\Delta \nu$ increases with the azimuthal angle until a value of $\sim$72~MHz/V. As shown in Fig.~\ref{fig:eom-sim}(a) and (c), the EO tuning efficiency increases with decreased waveguide-electrode spacing, reaching a value of $\sim$130~MHz/V for $d_{\text{w-e}} \sim 2~{\rm \mu m}$. However, our experiment testing shows that such a small $d_{\text{w-e}}$ would degrade the intrinsic optical Q of the resonator due to the perturbation of electrodes to the optical mode. To prevent such impact, we use a $d_{\text{w-e}}$ value of 4~${\rm \mu m}$ in the fabricated devices, which will leave the optical Q intact.

\subsection{On the mechanism of EO modulation of the soliton repetition rate}

The repetition rate of the Kerr solitons is determined by the free-spectral range (FSR) of the resonator at the carrier frequency of the solitons. Due to the group-velocity dispersion of the resonator, FSR is generally frequency dependent. As a result, a shift of the soliton carrrier frequency, $\delta \Omega$, would translate into a change of FSR, $\delta ({\rm FSR})$ given by the following expression
\begin{eqnarray}
\frac{\delta ({\rm FSR})}{{\rm FSR}} = - v_g \beta_2 \delta \Omega, \label{DFSR}
\end{eqnarray}
where $v_g$ and $\beta_2$ are the group velocity and group-velocity dispersion, respectively. The carrier frequency of soliton is impacted by the Raman-induced self-frequency shift, $\Omega_R$, which is related to the laser-cavity detuning of the pump wave, $\Delta$, as \cite{yi2016theory}
\begin{eqnarray}
\Delta = A \sqrt{\Omega_R} - B \Omega_R^2, \label{DetuningOmega_R}
\end{eqnarray}
where $\Delta=\omega_0-\omega_p$ is the frequency detuning between the cavity resonance $\omega_0$ and the pump laser frequency $\omega_p$. $A \equiv \sqrt{\frac{15 c |\beta_2| \omega_0}{32 n_0 Q \tau_R}}$ and $B \equiv \frac{c\beta_2}{2n_0}$ where $c$ is the velocity of light in vacuum. $n_0$ and $\tau_R$ are the refractive index and the Raman time constant of LN, respectively. $Q$ is the loaded optical Q of the device. A small change of the laser-cavity detuning, $\delta \Delta$, would lead to a small change of SFS of the soliton, $\delta \Omega_R$, given by
\begin{eqnarray}
\delta \Omega_R = \frac{\delta \Delta}{\frac{A}{2\sqrt{\Omega_R}}-2B\Omega_R}. \label{DOmegaR}
\end{eqnarray}

EO modulation of the comb resonator would result in time-dependent modulation of the cavity resonance $\delta \omega_0 (t) = \eta V_p \cos(\Omega_m t) = \delta \Delta (t)$, where $\eta$ is the EO tuning efficiency, $V_p$ is the peak value of the driving voltage, and $\Omega_m$ is the modulation frequency. Using this expression in Eq.~(\ref{DOmegaR}) and then apply it in Eq.~(\ref{DFSR}), we thus obtain the time-dependent modulation on the FSR given by
\begin{eqnarray}
\frac{\delta ({\rm FSR})}{{\rm FSR}} = \frac{- v_g \beta_2 \eta V_p \cos(\Omega_m t) }{\frac{A}{2\sqrt{\Omega_R}}-2B\Omega_R}. \label{DFSR_time}
\end{eqnarray}
As a result, the FM efficiency of the microwave, defined as the FSR change per unit driving voltage, is given by the following expression
\begin{eqnarray}
\rho_{\rm FM} = \frac{- v_g \beta_2 \eta {\rm FSR} }{\frac{A}{2\sqrt{\Omega_R}}-2B\Omega_R}. \label{FMEfficiency}
\end{eqnarray}

\begin{figure}[htpb]
	\centering\includegraphics[width=1\linewidth]{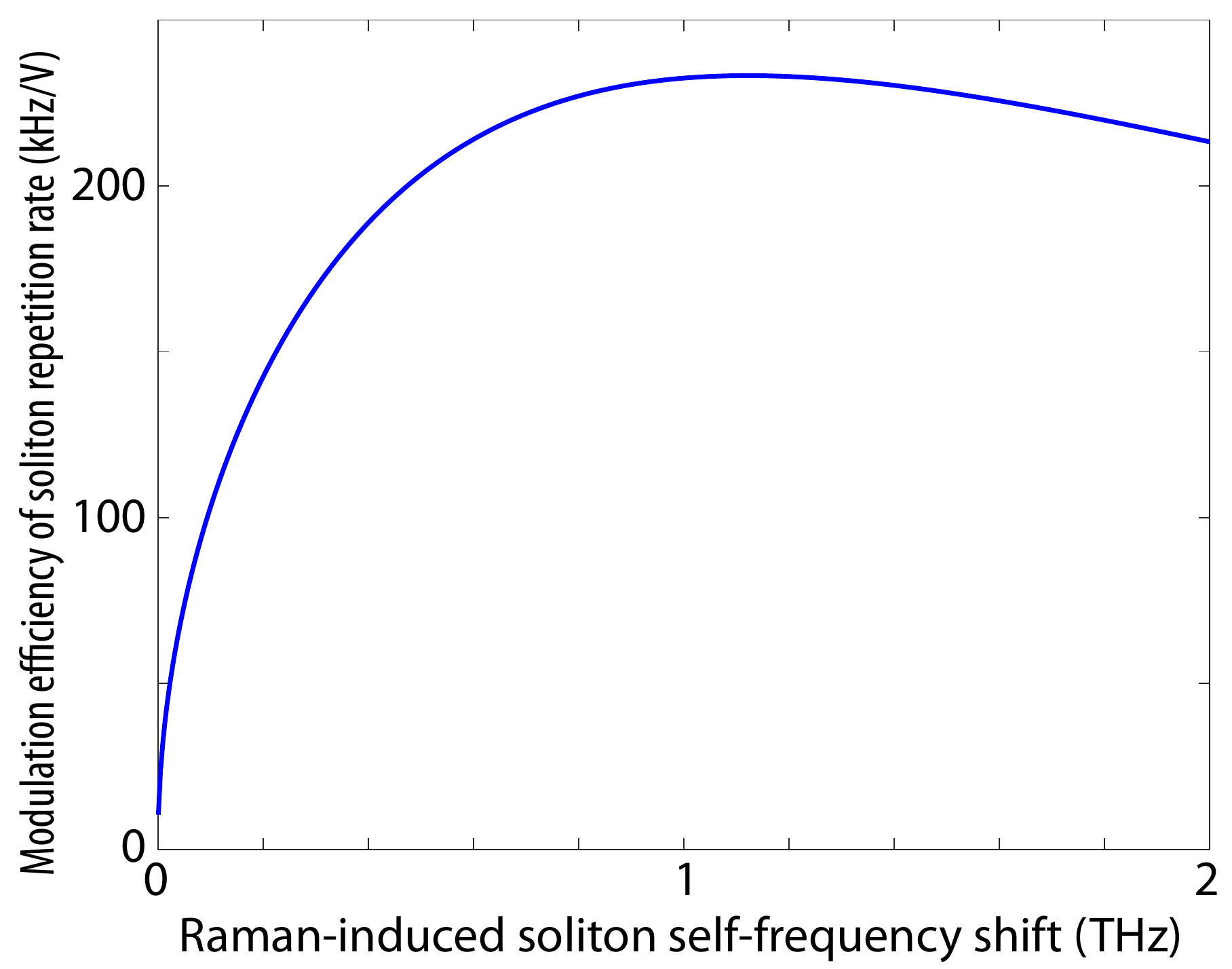}
	\caption{Theoretical FM efficiency as a function of Raman-induced SFS of soliton. The device parameters same as Fig.~\ref{Fig3} of the main text are used in the theoretical calculation, with the loaded optical Q of 3.2 million, group index $n_g = 2.357$, FSR = 19.81~GHz. $\tau_R = 6.3$~fs is adopted from Ref.~\cite{he2019self}. } \label{Fig_RepRate}
\end{figure}

Figure \ref{Fig_RepRate} shows the calculated theoretical FM efficiency of our device. Due to the spectral distortion induced by the coupling waveguide (Fig.~\ref{Fig2}), it is difficult to quantify the amount of SFS of the soliton combs. As an estimate, it is generally in the range of (1-2)~THz \cite{he2019self}. As shown in Fig.~\ref{Fig_RepRate}, the induced FM efficiency is roughly in the range of (100--200)~kHz/V. This level of FM efficiency explains well the observed phenomena at a low modulation frequency. However, it is significantly lower than we observed at a high modulation frequency of 50 and 75~MHz. The underlying mechanism is not clear at this moment and will require further exploration. 

\subsection{Estimation of FM amplitude at a high modulation frequency}

The electric spectrum analyzer we employed (Tektronix, RSA5126B) has a limited bandwidth of 160~MHz for time-dependent frequency characterization (although it does not have such bandwidth limitation when it is used for spectrum analysis). So for a high modulation frequency such as 75~MHz, it can only captures the first-order modulation sidebands. However, the spectrum of the microwave (Fig.~3(e) of the main text) shows multiple orders of modulation sidebands, implying that the spectrogram analysis likely underestimate the amplitude of frequency modulation (FM). Here we use the spectrum of the microwave for a rough estimate of the FM amplitude. 

The theory in the previous section shows that the modulation of soliton repetition rate results from the Raman-induced SFS via the modulation of soliton energy  \cite{yi2016theory}. As a result, it is expected that the frequency modulation (FM) of the detected microwave will be accompanied with a certain extent of amplitude modulation (AM). In the time domain, an electric field $E$ of a microwave that is undergoing time-varying phase modulation as well as amplitude modulation can be expressed as follows:
\begin{equation}
E(t) = \frac{E_0}{\sqrt{1+a^2}} [1 + a \sin(\Omega_m t)] e^{i\left(\Omega_0 t + \sigma \cos \Omega_m t\right)} ,
\label{eq:pmfield}
\end{equation}
where $E_0$ is the field amplitude, $\Omega_0$ is the carrier frequency of the microwave, and $\Omega_m$ is the modulation frequency. $\sigma$ and $a$ are the modulation depths of phase modulation and amplitude modulation, respectively. The microwave field shown in Eq.~(\ref{eq:pmfield}) exhibits an instantaneous frequency as 
\begin{equation}
\Omega(t) = \Omega_0 - \sigma \Omega_m \sin \Omega_m t ,
\label{eq:Freq}
\end{equation}
which varies with time in a sinusoidal fashion with an FM amplitude of $\sigma \Omega_m$. In Eqs.~(\ref{eq:pmfield}) and (\ref{eq:Freq}), we have assumed that the FM and AM are out of phase, which is expected from the Raman-induced SFS effect of the soliton: The higher the soliton energy (and thus the larger the amplitude of the detected microwave), the larger the induced SFS and thus the longer the carrier wavelength of the soliton. The longer the carrier wavelength, the lower the repetition rate (and thus the lower the microwave frequency) due to anomalous GVD of the device. 

\begin{figure}[t!]
	\centering
	\includegraphics[width=1\linewidth]{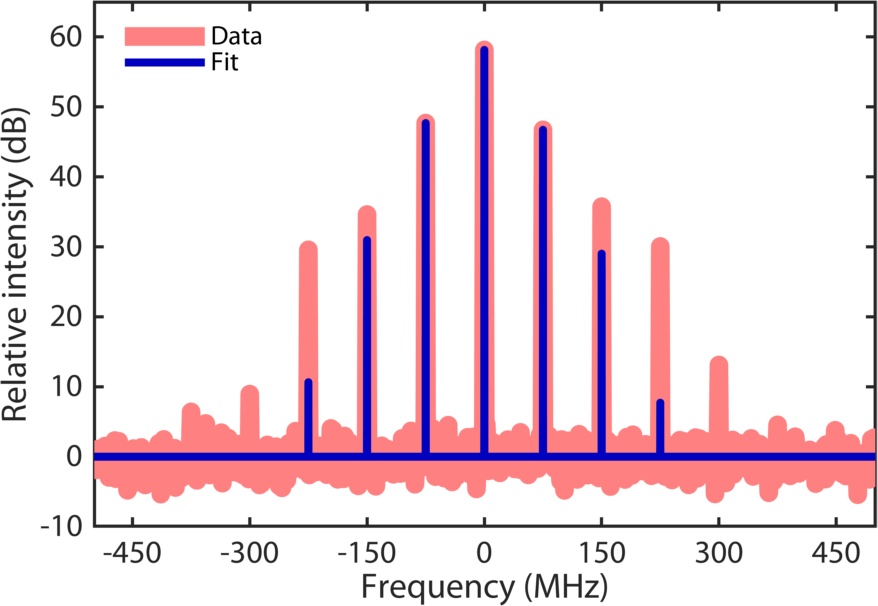}
	\caption[]{Comparison between the theoretical and experimentally observed spectra of the 19.81~GHz microwave with a modulation frequency of 75~MHz. Red line: experimentally recorded RF spectrum from Fig.~\ref{Fig3}(e) of the main text. Blue line: theoretically calculated spectrum, with $\sigma=0.55$ and $a=0.03$. }
	\label{fig:jacobi}
\end{figure}

Equation~(\ref{eq:pmfield}) can be expanded in terms of the Bessel functions as 
\begin{eqnarray}
E(t) &=& \frac{E_0 e^{i\Omega_0 t}}{\sqrt{1+a^2}} [1 + a \sin(\Omega_m t)] \sum_{n = -\infty}^{\infty} i^{n} J_n (\sigma) e^{i n \Omega_m t} \nonumber \\
&=& \frac{E_0 e^{i\Omega_0 t}}{\sqrt{1+a^2}} \sum_n {i^n e^{n \Omega_m t} \left[ J_n - \frac{a }{2} J_{n-1} - \frac{a }{2} J_{n+1} \right]  }, \nonumber\\
\label{eq:Bessel}
\end{eqnarray}
where $J_n(\sigma)$ is the $n^{th}$ order Bessel function of the first kind. High-speed phase modulation (frequency modulation) produces a series of modulation sidebands whose frequencies are separated from the carrier frequency by an integer number of the modulation frequency with amplitudes scaled with the Bessel function coefficients $J_n(\sigma)$ whose magnitude depends on the modulation depth $\sigma$. We can thus use this approach to provide a rough estimate on the modulation depth, by comparing the theoretical spectrum with the experimentally recorded one. Figure \ref{fig:jacobi} shows the results for the modulation frequency of 75~MHz. A phase-modulation depth of $\sigma=0.55$ and an amplitude modulation depth of $a=0.03$ provide a reasonably good description of the modulation sidebands up to the first two orders whose relative amplitudes agree closely the experimental ones. Accordingly, the FM amplitude, $\sigma\Omega_m/(2\pi)$, is estimated to be $\sim 41$~MHz. Figure \ref{fig:jacobi} shows certain discrepancies on the higher-order sidebands between the theoretical estimation and experiment data, which likely implies a certain distortion associated with the modulation. 

It is important to note that such a brief spectrum analysis can only provide a rough estimate of the level of frequency modulation, since the relative phases of the modulation sidebands play crucial roles on the real extent of FM, which cannot be obtained from the spectrum. The exact extent of FM/AM will require larger characterization bandwidth of the spectrum analyzer, which cannot be perform at this moment and will be left for future exploration.  

\medskip

\subsection{Modulation of soliton repetition rate at a speed beyond the photon lifetime limit of the comb resonator}

The driving electrodes of the device supports modulation at a speed beyond the photon lifetime limit of the resonator. At such a high speed, however, the employed spectrum analyzer cannot perform the time-dependent frequency analysis due to its bandwidth limit (160~MHz). However, it is still able to perform spectrum characterization of the microwave signal. Figure \ref{Figure5} shows two examples with a modulation frequency of 100 and 300~MHz, respectively. Clearly, since the modulation frequency is considerably beyond the photon lifetime limit of the comb resonator, the modulation efficiency drops considerably, as evident by the smaller amplitudes of the modulation sidebands, compared to the case of 75~MHz shown in the main text.

\begin{figure}[htpb]
	\centering\includegraphics[width=1\linewidth]{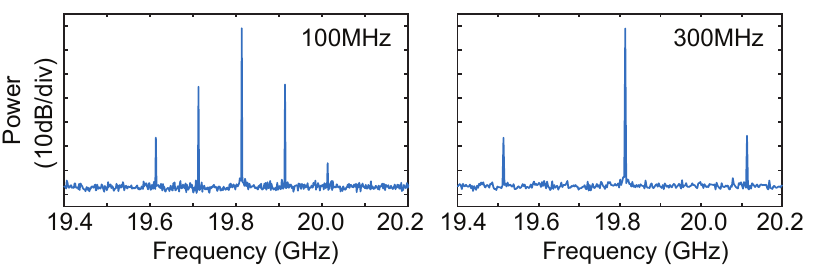}
	\caption{Microwave spectra for the modulation frequency at 100~MHz (left) and 300~MHz (right), with a peak driving voltage similar to the case of 75~MHz shown in the main text. } \label{Figure5}
\end{figure}


\if{
\subsection{Estimate of EO tuning efficiency at a high modulation frequency}

The electric spectrum analyzer we employed (Tektronix, RSA5126B) has a limited bandwidth of 160~MHz for time-frequency characterization (although it does not have such bandwidth limitation when it is used for spectrum analysis). So for a high modulation frequency such as 75~MHz, it can only captures the first-order modulation sidebands. However, the spectrum of the microwave (Fig.~3(e) of the main text) shows multiple orders of modulation sidebands, implying that the time-frequency analysis likely underestimate the amplitude of frequency modulation (FM). Here we use the spectrum of the microwave for a rough estimate of the FM amplitude. 

The theory in the previous section shows that the modulation of soliton repetition rate results from the Raman-induced SFS via the modulation of soliton energy  \cite{yi2016theory}. As a result, it is expected that the frequency modulation (FM) of the detected microwave will be accompanied with a certain extent of amplitude modulation (AM). In the time domain, an electric field $E$ of a microwave that is undergoing time-varying phase modulation and amplitude modulation can be expressed as follows:
\begin{equation}
E(t) = E_0 e^{i\left(\Omega_0 t + m \cos \Omega_m t\right)} ,
\label{eq:pmfield}
\end{equation}
where $E_0$ is the field amplitude, $\Omega_0$ is the carrier frequency of the microwave, $m$ is the modulation depth, and $\Omega_m$ is the modulation frequency. The microwave field shown in Eq.~(\ref{eq:pmfield}) exhibits an instantaneous frequency as 
\begin{equation}
\Omega(t) = \Omega_0 - m \Omega_m \sin \Omega_m t ,
\label{eq:Freq}
\end{equation}
which varies with time in a sinusoidal fashion with an FM amplitude of $m \Omega_m$.

\begin{figure}[b!]
	\centering
	\includegraphics[width=1\linewidth]{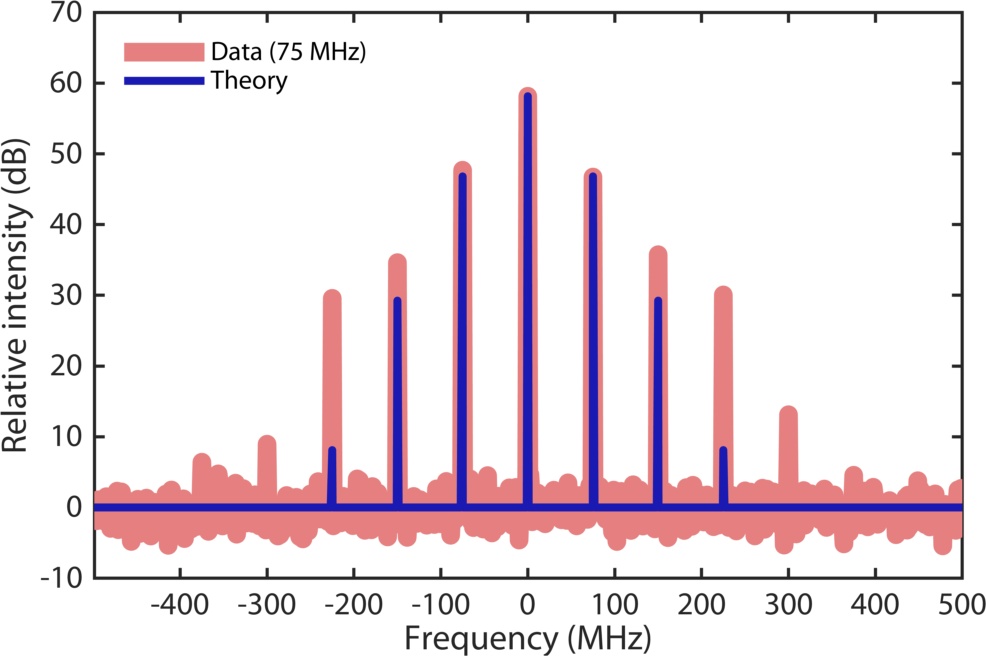}
	\includegraphics[width=1\linewidth]{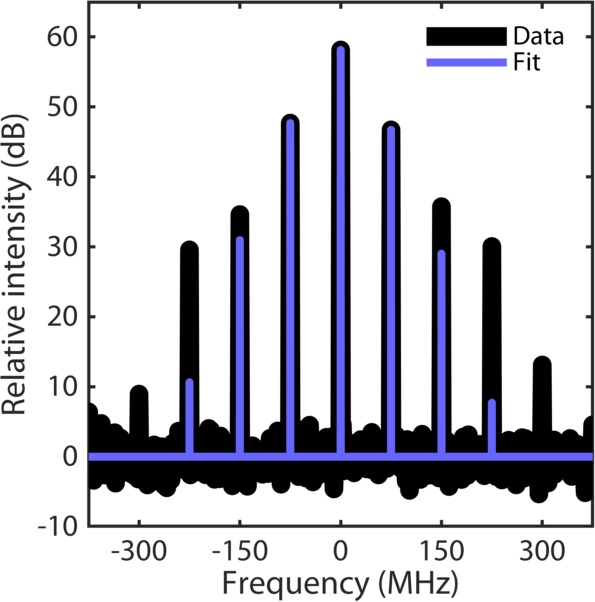}
	\caption[]{Comparison between the theoretical and experimentally observed spectra of the 19.81~GHz microwave with a modulation frequency of 75~MHz. Red line: experimentally recorded RF spectrum from Fig.~\ref{Fig3}(e) of the main text. Blue line: theoretically calculated spectrum used to estimate the modulation depth ($m = 0.52$) resulting from high-speed phase modulation.}
	\label{fig:jacobi}
\end{figure}

Via the Jacobi-Anger identity \cite{holmes1998general}, Eq.\ (\ref{eq:pmfield}) can be expanded in terms of the Bessel functions as 
\begin{equation}
E(t) = E_0 e^{i\omega t} \sum_{n = -\infty}^{\infty} i^{n} J_n (m) e^{i n \Omega t}
\end{equation}
where $J_n(\cdot)$ is the $n^{th}$ order Bessel function of the first kind. High-speed phase modulation (frequency modulation) produces a series of modulation sidebands whose frequencies are separated from the carrier frequency by an integer number of the modulation frequency with amplitudes scaled with the Bessel function coefficients $J_n(m)$ whose magnitude depends on the modulation depth. We can thus use this approach to provide a rough estimate on the modulation depth, by comparing the theoretical spectrum with the experimentally recorded one. Figure \ref{fig:jacobi} shows the results for the modulation frequency of 75~MHz. A modulation depth of $m=0.52$ provides the first two sidebands whose relative amplitudes agree closely the experimental ones. Therefore, the amplitude of frequency modulation, $m\Omega_m/(2\pi)$, is estimated to be about 6.2~MHz. The slight discrepancy between the theory and experiment data is likely due to the residual amplitude modulation of the soliton comb.

\subsection{Modulation of soliton repetition rate at a speed beyond the photon lifetime limit of the comb resonator}

The driving electrodes of the device supports modulation at a speed beyond the photon lifetime limit of the resonator. At such a high speed, however, the employed spectrum analyzer cannot perform the time-frequency analysis due to its bandwidth limit (160~MHz). However, it is still able to perform spectrum characterization of the microwave signal. Figure \ref{Figure5} shows two examples with a modulation frequency of 100 and 300~MHz, respectively. Clearly, since the modulation frequency is considerably beyond the photon lifetime limit of the comb resonator, the modulation efficiency drops considerably, as evident by the smaller amplitudes of the modulation sidebands, compared to the case of 75~MHz shown in the main text.

\begin{figure}[htpb]
	\centering\includegraphics[width=1\linewidth]{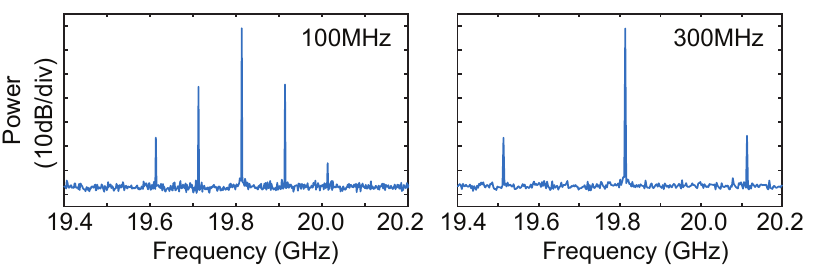}
	\caption{Microwave spectra for the modulation frequency at 100~MHz (left) and 300~MHz (right), with a peak driving voltage similar to the case of 75~MHz shown in the main text. } \label{Figure5}
\end{figure}

\subsection{Analysis of EO tuning efficiency and harmonic distortion}

\begin{equation}
A_0 = \frac{1}{\sqrt{1 + \sum_{n=1}^N a_n^2}}
\end{equation}

\begin{equation}
E_{\mathrm{AM}} = 1 + \sum_{n=1}^N a_n \cos \left[ n \left( \Omega_m t + \phi_0 \right) \right]
\end{equation}

\begin{equation}
E_{\mathrm{FM}} = \exp \left[ i \sum_{n=1}^N m_n \cos  \left( n \Omega_m t \right) \right]
\end{equation}

\begin{equation}
E = A_0 E_{\mathrm{AM}} E_{\mathrm{FM}} e^{i \omega_0 t}
\end{equation}

}\fi

\if{

In order to estimate the shift $\delta\omega_0$ of the angular resonance frequency $\omega_0$ induced via electro-optic modulation (EOM), we employ a perturbative analysis based on the Hellman-Feynman theorem \cite{johnson2002perturbation}:
\begin{equation}
\delta\omega_0 = - \frac{\omega_0}{2} \frac{ \left< E \right| (\delta\epsilon_r)_{ij} \left| E \right>}{ \left< E \right|  \hat{\epsilon}_r \left| E \right>},
\label{eq:deltaom0}
\end{equation}
where $\hat{\epsilon}_r$ is the relative permittivity tensor, $(\delta\epsilon_r)_{ij}$ is the EOM-induced perturbation to the latter, and $\ket{E}$ is the electric field within the microring resonator. 

To calculate $(\delta\epsilon_r)_{ij}$ induced by a driving electric field $\ket{\mathcal{E}}$, we perform some algebra using the impermeability tensor $\hat{\eta} \equiv \hat{\epsilon}_r^{-1}$, and we obtain a result using the Pockels tensor $\hat{r} = r_{ijk} \equiv (\partial \eta_{ij}/\partial \scE_k)_{\scE=0}$:
\begin{equation}
(\delta\epsilon_r)_{ij} = - \hat{\epsilon}_r (\hat{r} \ket{\scE} ) \hat{\epsilon}_r
= (\epsilon_r)_{ij} r_{jkl} \scE_l (\epsilon_r)_{ks}.
\label{eq:delep}
\end{equation}
If we take the lateral extent of the microring to be in the $xy$-plane, for $z$-cut Lithium niobate (LN) we use a diagonal matrix for $\hat{\epsilon}$ with the $z$-axis corresponding to the optic axis,
\begin{equation}
\hat{\epsilon} = 
\begin{pmatrix}
\epsilon _{11} & 0 & 0 \\
0 & \epsilon _{11} & 0 \\
0 & 0 & \epsilon _{33} \\
\end{pmatrix},
\end{equation}
where $\epsilon_{11}$ and $\epsilon_{33}$ are the relative permittivity coefficients. With the same coordinates, to calculate Eq.\ \ref{eq:delep} we use the contracted form $r_{ij}$ of $\hat{r}$,
\begin{equation}
	r_{ij} = 
\begin{pmatrix}
	0 & -r_{22} & r_{13} \\
	0 & r_{22} & r_{13} \\
	0 & 0 & r_{33} \\
	0 & r_{51} & 0 \\
	r_{51} & 0 & 0 \\
	-r_{22} & 0 & 0 
\end{pmatrix}.
\end{equation}

In order to obtain the most general result, we expand the expression in Eq.\ \ref{eq:deltaom0} and divide by $2\pi$ to obtain the absolute resonance frequency shift $\delta\nu_0$:
\begin{equation}
\delta\nu_0 = \sum_{u}\sum_{v} \delta\nu_u^{(v)},
\label{eq:dnu}
\end{equation}
where we have broken down the total integration into a sum of several Cartesian components $\delta\nu_u^{(v)}$:
\begin{equation}
\delta\nu_u^{(v)} = - \frac{\nu_0}{2} \frac{\int_{\text{core}} \Diff3\mathbf{r} \xi_u^{(v)}}{\int_{\text{all}}  \Diff3\mathbf{r} \bar{\xi} }, \quad u \in \{x, y, z\},\quad v \in \{1, 2\},
\end{equation}
where we integrate the perturbation term only within the LN core and normalize the optical field over all space. $\nu_0$ is the absolute resonance frequency, $\bar{\xi}$ is the normalization factor for the optical fields,
\begin{equation}
\bar{\xi} = \epsilon _{11} \left(\left| E_x\right|^2+\left| E_y\right|^2\right)+\epsilon _{33} \left| E_z\right|^2,
\end{equation}
and $\xi_u^{(v)}$ are integrands where the subscript $u$ corresponds to the driving field Cartesian component and the superscript $v$ is an arbitrary index,
\begin{equation}
\begin{split}
\xi_x^{(1)} &= 2 \mathcal{E}_x r_{51} \epsilon _{33} \epsilon _{11} \operatorname{Re}\left\{E_x E_z^*\right\}, \\
\xi_x^{(2)} &= -2 \mathcal{E}_x r_{22} \epsilon _{11}^2 \operatorname{Re}\left\{E_x E_y^*\right\}, \\
\xi_y^{(1)} &= \mathcal{E}_y r_{22} \epsilon _{11}^2 \left(\left| E_y\right|^2-\left| E_x\right|^2\right), \\
\xi_y^{(2)} &= 2 \mathcal{E}_y r_{51} \epsilon _{33} \epsilon _{11} \operatorname{Re}\left\{E_y E_z^*\right\}, \\
\xi_z^{(1)} &= \mathcal{E}_z  r_{13} \epsilon _{11}^2 \left(\left| E_x\right|^2+\left| E_y\right|^2\right), \\
\xi_z^{(2)} &= \mathcal{E}_z r_{33} \epsilon _{33}^2 \left| E_z\right|^2.
\end{split}
\end{equation}
The latter expressions are useful for distinguishing the contributions of various components from both the driving and optical electric fields. Also, it turns out that only $\xi_x^{(2)}$ and $\xi_y^{(1)}$ have any appreciable contribution since the electric field components $\scE_z$ and $E_z$ are close to null. This is the case because our $z$-cut device is designed to use the quasi-TE mode and our electrode pairs are positioned in the device plane.

\begin{figure}[htp!]
	\centering
	\includegraphics[width=1\linewidth]{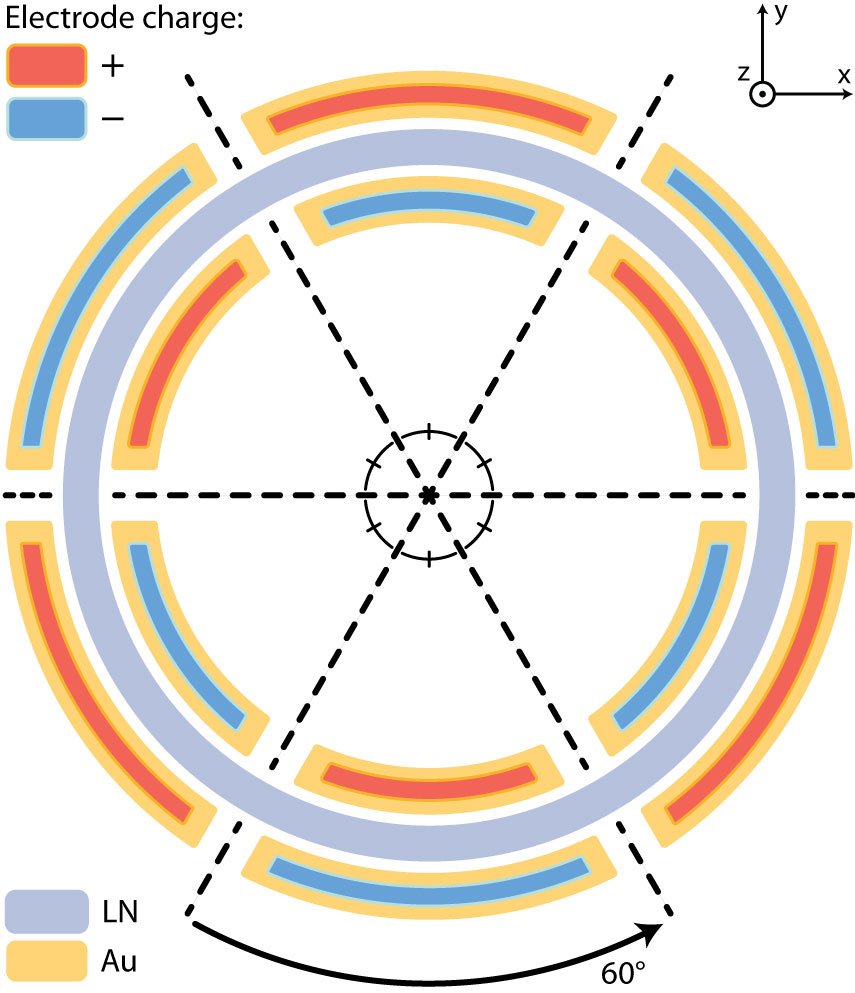}
	\caption[]{Layout for electro-optic modulation of $z$-cut microring resonator. There are six pairs of electrodes with alternating polarities all arranged symmetrically and spaced azimuthally by $60\deg$ along the ring circumference. Colors indicate Lithium niobate (LN), gold (Au), or electrode charge ($+$ or $-$).}
	\label{fig:zcut}
\end{figure} 

With the latter knowledge, we can obtain a rough estimate for the magnitude of $\delta\nu_0$ prior to any mode simulations. For the optical field $\ket{E}$ we define a TE mode in 2D polar coordinates $(\rho, \theta)$:
\begin{equation}
\ket{E} = E_0 \left( \cos\theta \hat{e}_x + \sin\theta \hat{e}_y \right)\delta(\rho - R),
\label{eq:eopt}
\end{equation}
where $E_0$ is the optical field amplitude, $R$ is the microring radius, $\delta(\cdot)$ is the Dirac delta function (which we assume since $R$ is much greater than the microring waveguide width), and $\hat{e}_x$  $(\hat{e}_y)$ is the unit vector in the $x$ ($y$) axis. Similarly, because the electrodes are placed on opposite sides of the ring waveguide, $\ket{\scE}$ is also in the transverse plane:
\begin{equation}
\ket{\scE} = \scE_0 \left( \cos\theta \hat{e}_x + \sin\theta \hat{e}_y \right)\delta(\rho - R),
\label{eq:eac}
\end{equation}
where $\scE_0$ is the driving field amplitude.

After plugging these fields into Eq.\ \ref{eq:dnu} and only integrating ($\Diff3\mathbf{r} \rightarrow \rho \diff\rho \diff\theta$) with respect to $\rho$ from $0$ to $\infty$, we see that the remaining $\theta$ integrand $\xi(\theta)$ in the numerator only depends on $r_{22}$ and $\epsilon_{11}$,
\begin{equation}
\xi(\theta) = \int_{\text{core}} \rho\diff\rho\left(\xi_x^{(2)} + \xi_y^{(1)} \right) = -r_{22} \epsilon_{11}^2 \sin 3\theta.
\label{eq:etath}
\end{equation}
Finally, integrating $\theta$ from $0$ to $\theta_0$, we obtain an oscillatory result,
\begin{equation}
\delta\nu_0 \Big|_0^{\theta_0} = \frac{1}{6} \scE_0 \nu_0 r_{22} \epsilon_{11} \left( 1 - \cos3\theta_0 \right),
\end{equation}
which peaks at $\max\{\delta\nu_0\} = \frac{1}{3} \scE_0 \nu_0 r_{22} \epsilon_{11}$ for $\theta = (2n + 1)\pi/3$ $(n \in \{0,1,2,\dots\})$ but yields a total of $0$ after setting $\theta_0 = 2\pi$, i.e.\ integrating over the entire ring yields zero net frequency shift. However, as shown in Fig.\ \ref{fig:zcut}, in order to improve and maximize the total shift $\delta\nu$, we can alternate the sign or polarity of the driving field across six identical electrode pairs placed consecutively along the microring circumference, modifying $\scE_0$ into a function of $\theta$. In this way, the maximum achievable value for $\delta\nu_0$ is newly calculated by replacing $\scE_0$ in Eqs.\ \ref{eq:eopt} and \ref{eq:eac} by a square wave $\scE_{\theta}$ with period $2\pi/3$ prior to integration, i.e.\
\begin{equation}
\scE_0 \rightarrow \scE_{\theta} = \scE_0 \operatorname{sign}\left(\sin 3 \theta\right),
\end{equation}
and consequently $\xi(\theta)$ is also modified into $\xi'(\theta)$:
\begin{equation}
\xi(\theta) \rightarrow \xi'(\theta) = -r_{22} \epsilon_{11}^2 \sin 3\theta \operatorname{sign}\left(\sin 3 \theta\right). 
\end{equation}
This yields the optimal result $\Delta\nu$, whereby the $\delta\nu_0$ maximum is enhanced by a factor of six after integrating Eq.\ \ref{eq:etath} from 0 to $2\pi$ radians:
\begin{equation}
\Delta\nu = 6 \max\{\delta\nu_0\} = 2 \scE_0 \nu_0 r_{22} \epsilon_{11}.
\end{equation}

With the latter expression we can now estimate the total shift we should expect for our device with either just one electrode (as we have managed in our devices so far) or six electrodes, yielding 0.088 and 0.53 GHz/V, respectively. Here we set $\epsilon_{11}$, $r_{22}$, $\nu_0$, and $\scE_0$ to $2.21^2$, $7$ pm/V, $193.4$ THz, and $0.04$ V/$\mu m$, respectively, and the $\scE_0$ value was obtained from an electric field simulated within the LN waveguide core using COMSOL (Fig.\ \ref{fig:comsol-mode}).

\begin{figure}[htbp!]
	\centering
	\includegraphics[width=1\linewidth]{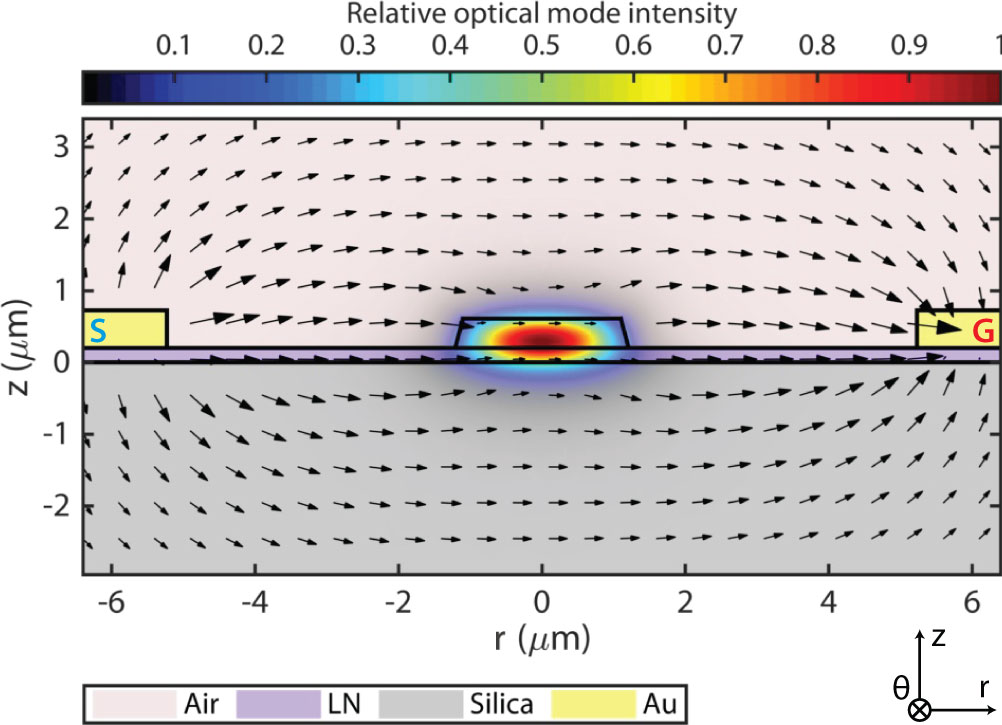}
	\caption[]{LN microring quasi-TE$_{00}$ optical mode profile and vector map of applied electric field. Device materials are indicated in different colors: yellow, lavender, gray, and white correspond to gold, LN, silica, and air, respectively. The black arrows represent the real components of both $\scE_x$ and $\scE_y$, respectively. The colormap on top represents the normalized relative intensity of the optical mode profile.}
	\label{fig:comsol-mode}
\end{figure} 

In order to solidify our analysis, we carried out detailed COMSOL simulations where the optical and applied electric field vector components were used to calculate the maximum shift $\Delta\nu$. The results are presented in Fig.\ \ref{fig:eom-sim}, which show the dependence of $\Delta\nu$ on both the electrode coverage angle $\theta$ and the waveguide-electrode spacing $d_{\text{w-e}}$. It is evident that when we incorporate the electrode scheme from Fig.\ \ref{fig:zcut}, $\Delta\nu$ grows with angle in a manner reminiscent of a quasi-phase matched nonlinear process. Because we utilize just one electrode pair for modulation, our tuning slope corresponds to an angle of about 1.04 rad, and, for $d_{\text{w-e}} \sim 4$ $\mu m$, from the middle plot in Fig.\ \ref{fig:eom-sim} we gather that this is approximately 0.077 GHz/V, a value only $\sim 10$ percent less than the one we calculated previously by assuming the fields had scalar amplitudes. Further, from the bottom plot in Fig.\ \ref{fig:eom-sim}, we see that at our operating wavelength of 1.55 $\mu m$, if the whole of $2\pi$ radians are modulated with our alternating electrode scheme, with $d_{\text{w-e}} \sim 2$ $\mu m$ we can achieve up to $\sim 0.8$ GHz/V tuning rates, which corresponds to $\sim 6.4$ pm/V.

\begin{figure}[hbp!]
	\centering
	\includegraphics[width=1\linewidth]{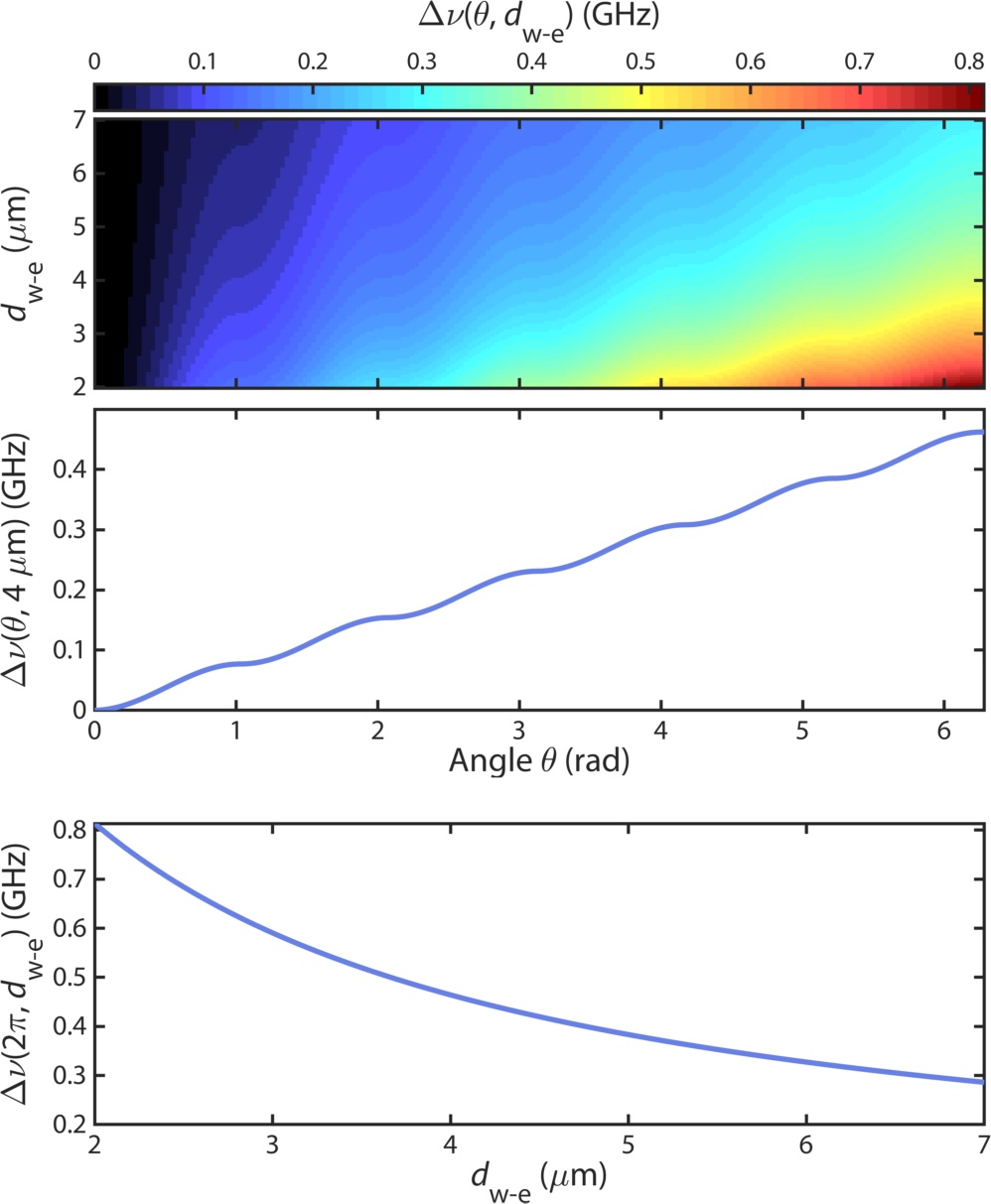}
	\caption[]{Simulated frequency shift $\Delta\nu$ vs.\ azimuthal angle $\theta$ and waveguide-electrode spacing $d_{\text{w-e}}$ for a resonance frequency of $\sim$ 193.4 THz (1.55 $\mu m$). Top: surface plot vs.\ $\theta$ and $d_{\text{w-e}}$ where the colormap represents the frequency shift $\Delta\nu$. Middle: $\Delta\nu$ vs.\ electrode coverage angle $\theta$ for $d_{\text{w-e}} \sim 4$ $\mu m$. Bottom: $\Delta\nu$ vs.\ waveguide-electrode spacing $d_{\text{w-e}}$ with a full $2\pi$ microring electrode coverage.}
	\label{fig:eom-sim}
\end{figure} 

}\fi

\end{document}